\documentstyle[10pt]{article}
\begin{document}
\begin{titlepage}
%
\begin{flushright}
{\bf IFUM 624/FT} \\
{\bf MPI/Pht 98-39}
\end{flushright} 
\begin{center}

\setlength {\baselineskip}{0.3in}
{\Large\bf Constructive algebraic renormalization
\\ of the abelian Higgs-Kibble model}
\vspace{1cm} \\
\setlength {\baselineskip}{0.2in}

{\large Ruggero Ferrari\footnote{ruggero.ferrari@mi.infn.it}}

\vspace{0.1in}
{\it Dipartimento di Fisica, Universit\`a di Milano\\
via Celoria 16, 20133 Milano, Italy \\
and \\ INFN, Sezione di Milano}
\vspace{0.1in}
\par
{\large and}
\vspace{0.1in}
\\
\setlength {\baselineskip}{0.2in}

{\large Pietro Antonio Grassi\footnote{pgrassi@mppmu.mpg.de}}

\vspace{0.1in}
{\it Max-Planck-Institut f\"ur Physik\\
(Werner-Heisenberg-Institut)\\
F\"ohringer Ring 6, 80805 M\"unchen, Germany}
\vspace{1cm}\\
\setlength {\baselineskip}{0.2in}
{\bf Abstract \\}
\end{center}
We propose an algorithm, based on Algebraic Renormalization,
that allows the restoration of Slavnov-Taylor invariance
at every order of perturbation expansion for an anomaly-free
BRS invariant gauge theory. The counterterms are explicitly constructed
in terms of a  set of one-particle-irreducible 
Feynman amplitudes evaluated at zero momentum (and derivatives
of them).  The approach is here discussed in the case  of
the abelian Higgs-Kibble model, where the zero momentum
limit can be safely performed. 
The normalization conditions are imposed by means of the
Slavnov-Taylor invariants and are chosen in order to simplify the
calculation of the counterterms. In particular within
this model all counterterms involving BRS external sources
(anti-fields) can be put to zero with the exception of the
fermion sector. 
\setlength{\baselineskip}{0.3in}         
\end{titlepage}
%
\section{Introduction}   
\label{sec:intr}
Few gauge models of physical interest enjoy a symmetrical 
regularization of Feynman amplitudes (as QCD in dimensional regularization). 
In particular for the standard model the difficulty comes from 
the endemic presence of $\gamma_5$ and of the complete 
antisymmetric tensor. 
Thus, if the regularization breaks the desired symmetries, 
one has to recover the correct Green's functions 
by finite renormalization in order to fulfill the  
Slavnov-Taylor Identities (STI). Algebraic Renormalization (AR) 
(\cite{becHK}, \cite{becI}, \cite{stora}, \cite{algeb} and \cite{pig}) 
theory gives the conditions under which this strategy is possible: in particular
there should be no anomalies in the STI. Thus
in principle the renormalization program can be performed.
However it is not an easy task beyond
the one-loop approximation, since a high number of vertex functions 
at lower order must be evaluated for generic external momenta
in order to restore the STI. 
\par
In this paper we propose a strategy for the evaluation
of the counterterms, based on the zero-momentum
subtraction \cite{zim}. The final result is an explicit solution
of the STI where the counterterms are given in terms
of a set of finite vertex functions and their derivatives
evaluated at zero momentum.
Our strategy  is based on various results
taken from BPHZL renormalization scheme and from the Algebraic
Renormalization theory. We show that the zero momentum
subtraction and a judicious use of the normalization
conditions allows a practical evaluation of the counterterms 
by means of a relevant set of finite  vertex functions.
In particular the choice of the normalization
conditions entails a diagonal block structure
of the matrices that fix the counterterms. 
\par
As a starting point we assume that a consistent
subtraction procedure allows the evaluation of the n-loops vertex functions 
$\Gamma^{(n)}$ when the correct vertex functional ${{\rm I}\!\Gamma}^{j}$ is
given for any $j<n$. I.e. we assume that our procedure has
successfully worked for the lower orders and we proceed to restore
ST invariance on $\Gamma^{(n)}$. The n-order vertex functions
are constructed by iterative use of subgraphs and counterterms 
according to the scheme of Bogoliubov\cite{bogo}.
The regularization can
be any, provided  it respects the Quantum
Action Principle \cite{QAP} (i.e. it is correct up to counterterms in the action).
In order to make the discussion simpler we assume also that the
regularization procedure respects some basic symmetries of the
classical action, as Lorentz covariance, Faddeev-Popov (FP) charge
conservations and any possible further symmetry 
(as charge conservation C). Then we
expect that  Slavnov-Taylor identities (STI) are broken
\begin{eqnarray}
S(\Gamma)^{(n)}= \int d^4 x \left[\partial^\mu c \Gamma^{(n)}_{A^\mu}
+(\partial^\mu A_\mu + {{ev}\over{\alpha}}\phi_2)
\Gamma_{\bar c}^{(n)}\right] + \left(\Gamma,\Gamma\right)^{(n)}
=\Delta^{(n)}
\label{intr.1}
\end{eqnarray}
where the bracket is
\begin{eqnarray}
\left(X,Y \right) = {{\delta X}\over{\delta J_1}}{{\delta
Y}\over{\delta\phi_1}} 
+ {{\delta X}\over{\delta J_2}}{{\delta
Y}\over{\delta\phi_2}}  
-{{\delta X}\over{\delta \psi}}{{\delta
Y}\over{\delta{\bar\eta}}} 
+{{\delta X}\over{\delta {\bar\psi}}}{{\delta
Y}\over{\delta\eta}}
\label{intr.1p}
\end{eqnarray}
$J_1,J_1,\eta, {\bar\eta}$ are the sources coupled to the
BRS variations, i.e. the anti-fields (see the eq. (\ref{41a})).
We use the convention that derivatives are always from left and
any field with fermi character anti-commute. Although STI are broken,
the QAP guarantees that at every order 
$\Delta^{(n)}$ is a local insertion (provided that STI are valid
at the lower orders), has
the correct invariance properties under exact symmetries (e.g.
Lorentz, Faddeev-Popov (FP) charge, etc.) and
it is consistent with the power counting. Thus we can
expand $\Delta$ on a suitable basis
\begin{eqnarray}
\Delta = \sum_i c_i {\cal M}_i= \sum_i c_i\int d^4x f_i(\phi,\partial\phi)(x)
\label{intr.1.1}
\end{eqnarray}
where ${\cal M}_i$ is any Lorentz scalar monomial $f_i$
in the fields and their derivatives,
integrated over the Minkowski space. Residual symmetries
restrict the basis of ${\cal M}_i$; 
for the present Higgs-Kibble model 
C invariance constrains the possible breaking-terms. 
By construction the canonical dimension of ${\cal M}_i$
is less or equal five.
\par
The renormalization of the model consists in finding the finite
counterterms in the action that restore the validity of the STI 
and consequently the physical unitarity (Algebraic Renormalization).
Let us denote by ${{\rm I}\!\Gamma}$ the vertex functions resulting from this
procedure.
The locality and covariance of $\Delta$
suggests to consider the Taylor expansion in momentum
space. Let $t^{\delta}$ be the projector of the polynomials of
degree $\delta$ (the Taylor expansion in the independent
external momenta up to degree $\delta$) and $\delta_{\rm pc}$
denotes  the superficial degree of some given amplitude.
\par
These facts
suggest a strategy in the evaluation of the counterterms.
The first step consists in the zero momentum subtraction 
compatible with the power counting
\begin{eqnarray}
(1-t^{\delta_{\rm pc}})\Gamma.
\label{intr.2}
\end{eqnarray}
The above expression $t^\delta\Gamma$ is a short-hand notation of the following
procedure: one considers  first the relevant amplitude
(the functional derivatives respect to fields are denoted
by subscripts)
\begin{eqnarray}
\left. \Gamma_{\phi_1(p_1)\phi_2(p_2)...\phi_m(p_m)}
\right |_{p_m=-\sum_{j=1}^{m-1}p_j}
\label{intr.2.1}
\end{eqnarray}
then  the Taylor expansion $t^{\delta}$
in the independent momenta up to degree $\delta$. Formally
one has
\begin{eqnarray}
t^\delta\Gamma = \sum_{m=1}^{\infty} \int \prod_{i=1}^{m} d^4 p_i 
\phi_i (p_i) \delta^4 ({\scriptstyle \sum_{j=1}^{m}} p_j) t^{\delta}
\left .
\Gamma_{\phi_1(p_1)\phi_2(p_2)...\phi_m(p_m)}
\right |_{p_m=-\sum_{j=1}^{m-1}p_j}
\label{intr.2.2}
\end{eqnarray}
Thus we consider the lowest $\delta_{\rm D}$ such that
\begin{eqnarray}
(1-t^{\delta_{\rm D}})S(\Gamma)^{(n)}=
(1-t^{\delta_{\rm D}})\Delta^{(n)}=0
\label{intr.3}
\end{eqnarray}
at every order in the perturbation expansion ( $\delta_{\rm D}$
does not depend on $n$). Also the  
expression $t^{\delta_{\rm D}} \Delta^{(n)}$ has to be 
intended in the above sense and 
$\delta_D = 5 - \sum_{i} d_{\phi_i}$ ( $d_{\phi_i}$ are the
na\"\i ve dimensions of the fields, entering in the
functional derivatives of $\Delta^{(n)}$).
In the above equation the relevant
term for a recursive construction of the invariant vertex functions
is the linear operator on $\Gamma^{(n)}$
\begin{eqnarray}
S_0(\Gamma)^{(n)} \equiv 
\int d^4 x \left[\partial^\mu c \Gamma^{(n)}_{A^\mu}
+(\partial^\mu A_\mu + {{ev}\over{\alpha}}\phi_2)
\Gamma_{\bar c}^{(n)}\right] + \left(\Gamma^{(0)},\Gamma^{(n)}\right)
+ \left(\Gamma^{(n)},\Gamma^{(0)}\right)
\label{intr.4}
\end{eqnarray}
where $S_0$ is the linearized ST operator. 
We assume that zero momentum
subtraction is possible and focus our attention on other
effects of the subtraction. 
In general
$S_0$ is not homogeneous in the dimensions of the fields
(e.g. in presence of a spontaneous breaking of symmetry).
As a consequence the action of $(1-t^{\delta_D})$ on each 
single terms of $S_0(\Gamma^{(n)})$ induces some 
over-subtractions of $\Gamma^{(n)}$. These over-subtractions 
manifest theirselves as local breaking terms $\Delta$,
as can be seen by re-shuffling the 
eq. (\ref{intr.3}) in the form
\begin{eqnarray}
(1-t^{\delta_{\rm D}})S_0(\Gamma^{(n)})=
S_0((1-t^{\delta_{\rm pc}})\Gamma^{(n)})+
S_0(t^{\delta_{\rm pc}}\Gamma^{(n)})-
t^{\delta_{\rm D}}S_0(\Gamma^{(n)})
\label{intr.5}
\end{eqnarray}
The last terms show that the zero momentum subtraction does
not give ST invariant vertex functions and that order-by-order
we have to introduce counterterms in the action. Let us make
explicit the STI. The recursive
procedure gives
\begin{eqnarray}
&& S_0((1-t^{\delta_{\rm pc}})\Gamma^{(n)})
+\sum_{j=1}^{n-1}\left(
{{\rm I}\!\Gamma}^{(j)},{{\rm I}\!\Gamma}^{(n-j)}
\right)
\nonumber\\
&&=  \big[ t^{\delta_{\rm D}} S_0 - S_0t^{\delta_{\rm pc}}\big]
\Gamma^{(n)} +
t^{\delta_{\rm D}}\sum_{j=1}^{n-1}\left(
{{\rm I}\!\Gamma}^{(j)},{{\rm I}\!\Gamma}^{(n-j)}
\right) \equiv \Psi^{(n)}.
\label{intr.7}
\end{eqnarray}
The ${{\rm I}\!\Gamma}$ terms are computed at the lower
orders in the perturbative expansion.
They are supposed to satisfy STI at every order less than $n$. In our
strategy one of the criterion in the choice of the normalization conditions 
is the suppression of the above bilinear contributions.
\par
If the model has no anomalies the problem is then to find the
counterterms $\Xi^{(n)}$ which satisfy
\begin{eqnarray}
S_0(\Xi^{(n)}) = - \Psi^{(n)} 
\label{intr.8}
\end{eqnarray}
or
\begin{eqnarray}
S_0(\Xi^{(n)}) = - \big[ t^{\delta_{\rm D}} S_0 - S_0t^{\delta_{\rm pc}}\big]
\Gamma^{(n)} -
t^{\delta_{\rm D}}\sum_{j=1}^{n-1}\left(
{{\rm I}\!\Gamma}^{(j)},{{\rm I}\!\Gamma}^{(n-j)}
\right).
\label{intr.8a}
\end{eqnarray}
Finally the correct vertex functions are
\begin{eqnarray}
{{\rm I}\!\Gamma}^{(n)} = (1-t^{\delta_{\rm pc}})\Gamma^{(n)} + \Xi^{(n)}
\label{intr.9}
\end{eqnarray}
\par
The zero momentum subtraction, as intermediate renormalization, has
the advantage to reduce the renormalization in any subtraction 
procedure to a common ground: the algorithm is then the same
and it consists in the evaluation of a set
of finite amplitudes and their derivatives at zero momenta.
Moreover, as we will discuss later, it suggests a natural choice
of the  normalization conditions. Finally in the zero momentum
subtraction the contributions of the lower orders of perturbation
to $\Psi$ is consistently reduced (eq. (\ref{intr.7})).
\par
Among the problems of this approach is that the vertex functions
and their derivatives with respect external momenta must 
have regular behavior at zero momenta. In the 
presence of massless and massive fields, this requirement
implies the introduction of infra-red cut-offs and 
the Taylor operator $t^{\delta_D}$ has to be modified 
(see \cite{lowen}); however this
possibility will not be explored in the present work.
\par
There is a fairly large amount of freedom in the choice of
the counterterms $\Xi$ (eq. (\ref{intr.9})). This is due
to the presence of a certain number of ST invariant terms
explicitly given in the Appendix (\ref{sec:STI}).
This freedom will be exploited in order to obtain the most
efficient strategy in the evaluation of $\Xi$ and in order
to reduce the contribution to $\Psi$ (eq. (\ref{intr.8a})) due
to the lower perturbative terms. Any choice of
$\Xi$ fixes automatically the normalization conditions.
\par
The use of ST invariants and the normalization conditions
is organized by introducing a hierarchy for the counterterms
(choice of a basis of non-invariant counterterms).
They will be grouped into disjoint sets: the $S_0$ of two different
sets have no common elements. Subsequently the elements
of a single set can be organized with a nesting structure.
By following this hierarchy
decomposition, in the present model it is possible to
avoid all counterterms involving the external sources $J_i$,
tadpoles  and out-of-diagonal bilinear expressions. As a consequence
the mass counterterms turn out to be zero.
The ghost equation, which guarantees the nilpotency
of the ST operator, plays also an important r\^ole in the
control of some of the counterterms.
\par
By construction the functional $\Psi$ contains only finite
vertex functions, i.e. at every order of the perturbative 
expansion $n$ it can be evaluated independently from the regularization
procedure (once ${{\rm I}\!\Gamma}^{(j<n)}$ is correctly constructed). 
The counterterm functional $\Xi$ is determined
by eq. (\ref{intr.8}). In general there are more equations
than unknowns (over-determined problem). However the system
of equations has a solution since there are consistency conditions
\cite{wess}.
Most of them are consequence of the nilpotency of the ST operator
\begin{eqnarray}
S_0(\Psi^{(n)}) = 0.
\label{intr.10}
\end{eqnarray}
The evaluation of $\Xi$ can be performed either by imposing
the consistency conditions on $\Psi^{(n)}$ or by a choice of
the linearly independent equations. It should be remarked
that the expression of $\Xi^{(n)}$ in terms of $\Psi^{(n)}$
is a simple linear relation independent from the order of
the perturbation expansion.
\par
The really hard work is the evaluation of $\Psi^{(n)}$.
It consists in the computation of vertex functions and
of some of their derivatives at zero momenta.
The number of graphs turns out to be very large
(especially for amplitudes involving scalars). For this
reasons it is important to find possible relations among
the amplitudes, e.g. Callan-Symanzik Equation (CSE), and to
use automatic calculus to generate and evaluate the graphs.
Particularly interesting is the CSE (see for example 
\cite{bec_73}, \cite{bec_75}, \cite{krauss} and \cite{elis}).
The consistency conditions imposed by the CSE on the breaking terms
$\Psi^{(n)}$ allows the evaluation of some amplitudes in terms
of simpler vertex functions. Moreover some amplitude
can be obtained as the result of mass insertions on
vertex functions with less external legs. The automatic
calculus is particularly useful since the external momenta
are zero.
\par
It is important to reduce the contributions to $\Psi^{(n)}$
of the lower terms in the perturbation expansion. Eq.
(\ref{intr.7}) allows the direct control of the consequences
of any particular choice for the basis of the non-invariant
counterterms, i.e. of the choice of the normalization conditions.
This point of view is at variance with the on-shell conditions,
which cannot dispose this particular problem. For instance it
is clear that by dropping external sources counterterms
one can eliminate most of the terms coming from the lower
order in the perturbation expansion.
\par
The physical amplitudes necessitate the study of the zeros
of the two-point-functions. Then the free parameters of the
action have to be tuned in order to obtain the physical masses and the
correct coupling constants.
\par
The present paper is devoted to the $U(1)$ abelian Higgs-Kibble model 
(\cite{becHK}, \cite{pig_higg} )
for reasons of simplicity. The model has the advantage of admitting
dimensional regularization (if there is no fermion sector). 
It is non trivial, since
the presence of $\gamma_5$ requires the full generality of the
Algebraic Renormalization. Moreover the model has no anomalies: the
Adler-Bardeen-Jackiw anomaly is zero due to C-conjugation.
\par
In the following to make the formalism simpler and more direct 
we use to give a compact notation for the breaking-terms $\Psi^{(n)}$ 
and its coefficients: 
\begin{eqnarray}
\label{not}
\Psi^{(n)}=\sum_i\psi^{(n)}_{{\cal M}_i}{\cal M}_i
\end{eqnarray}
and in the same way we will denote the counterterms $\Xi^{(n)}$ 
by
\begin{eqnarray}
\label{not.1}
\Xi^{(n)}=\sum_k\xi^{(n)}_{{\cal P}_k}{\cal P}_k
\end{eqnarray}
where 
${\cal P}_i$ is a single monomial with dimension less or equal to four and 
null Faddeev-Popov charge. We may omit also the sign of integral
$\int d^4x$, when not necessary, e.g. $\xi^{(n)}_{\phi_1^2}
\equiv\xi^{(n)}_{\int d^4x\phi_1^2}$
\par
Eq. (\ref{intr.9}) can be looked from the point of view of a
different renormalization scheme. Let $\Gamma^{(n)}$ be the result
of any (non-symmetric) renormalization. One needs to introduce a set
of counterterms $\Gamma^{(n)}_{\rm CT}$ order-by-order:
\begin{eqnarray}
\Gamma^{(n)}+\Gamma^{(n)}_{\rm CT}
\label{stra.14}
\end{eqnarray}
By comparing with our procedure we have
\begin{eqnarray}
\Gamma^{(n)}_{\rm CT} = -t^{\delta_{\rm pc}}\Gamma^{(n)} + \Xi^{(n)}
+ \sum_j v_j {\cal I}_j.
\label{stra.15}
\end{eqnarray}
The first term is just a Taylor expansion of the action-like
amplitudes. The second term is evaluated in terms of finite amplitudes
and of some of their derivatives all at momentum zero. The later 
computation can be easily performed by automatic calculus. The
last term contains the ST invariants and accounts for the differences between
the normalization conditions in the two schemes.
\par
Section \ref{sec:hier} is devoted to the separation of the
counterterms into sectors. By a judicious choice of the
normalization conditions we can drop the tadpole and most of the
external source counterterms. Only in the fermion sector
the external source terms are modified by the renormalization
procedure. Moreover we can identify a bosonic, a kinetic-gauge
sector and a fermionic sector.
\par
Section \ref{sec:bterm} contains a study of the breaking term
functional $\Psi$. In particular the ST linearized operator
$S_0$ of eq. (\ref{intr.4}), which enters in expression
for $\Psi$, is modified in order to keep track of the ghost equation.
\par
Section \ref{sec:sol} provides the complete list of the counterterms
in terms of finite amplitudes. The solution contains the contribution
of the lower terms of the perturbative expansion. Moreover some
consistency conditions are shown to be present among the finite
amplitudes.
\par
Technical detail are in the Appendices. In Appendix \ref{sec:class}
we give the essential elements of the BRS transformations and
of the model. In Appendix \ref{sec:ST} we list all possible
counterterms and their ST transforms. In Appendix \ref{sec:STI}
we discuss the important issue of the linearly independent
ST invariants. Finally Appendix \ref{sec:psi} contains all the
relevant functional derivatives of the breaking term $\Psi$.
The expansion of the functional $\Psi$ in terms of Lorentz
invariant amplitudes allows the evaluation of the solutions
given in Section \ref{sec:sol}.

%
%
%
%
\section{Hierarchy of counterterms and breaking terms}
\label{sec:hier}
The complexity of the problem is somehow distributed
on two different 
steps. The evaluation of the breaking-terms
functional $\Psi$ is probably the most complex part. Once
$\Psi$ is given, one has to evaluate the counterterms
$\Xi$ by eq. (\ref{intr.8}). The present section is devoted
to this last problem.
In order to reduce the problem of managing 
the complete set of STI simultaneously, we introduce  
a hierarchy for the counterterms $\Xi$ and breaking terms $\Psi$. 
This problem has been already discussed in previous works (see \cite{becHK} and 
\cite{algeb}) on algebraic renormalization. 
\par
$S_0$ is a mapping of ${\cal V}_{\Xi}$ on ${\cal V}_{\Psi}$ 
\begin{eqnarray}
S_0 : {\cal V}_{\Xi} \to {\cal V}_{\Psi}
\label{hier.1}
\end{eqnarray}
where the vector spaces are given by the relevant
monomials 
\begin{eqnarray}
{\cal V}_\Xi\equiv \left\{\sum_k x_k
{\cal P}_k| x_k\in {\cal C},
{\rm dim({\cal P}_k)\leq 4, ~ FP~charge({\cal P}_k )=0}
\right\}
\label{hier.2}
\end{eqnarray}
and
\begin{eqnarray}
{\cal V}_\Psi\equiv \left\{\sum_i x_i
{\cal M}_i |  x_i\in {\cal C},
{\rm dim({\cal M}_i)  \leq 5, ~ FP~charge({\cal M}_i) =1}
\right\}.
\label{hier.3}
\end{eqnarray}
\par
The set of all action-like functionals $\{{\cal I}_i\}$
which are invariant
under ST transformations form the kernel of $S_0$
\begin{eqnarray}
\ker(S_0) =  \left\{\sum_i v_i
{\cal I}_i |  v_i\in {\cal C},{\rm dim({\cal I}_i)\leq 4,
~ FP~charge({\cal I}_i)=0}
\right\}.
\label{hier.4}
\end{eqnarray}
Some of the ST invariants are genuine BRS invariants.
The trivial ST invariants are given by all elements
which are $S_0$-variation of local functionals of dimension 
$\leq 3$ and FP charge = -1. The subspace
$\ker(S_0)$ induces an equivalence relation among the
counterterms.
The freedom of the choice of the
representative of the equivalence classes will be
used as one of the tools to organize the counterterms
in a hierarchy, according to a strategy aiming to reduce
the complexity of AR. This choice amounts to fix
the normalization conditions; in fact in this way we select
a basis on which we write the counterterm functional
$\Xi$. Therefore all monomials outside the basis do not
appear as counterterms. It should be mentioned here
that the sub-space ${\rm ker}(S_0)$ is further restricted
by the condition imposed by the ghost equation of motion.
The necessity to impose this condition as a first step
comes from the fact that the ghost equation of motion
is the statement of the nilpotency of $S_0$.
\par
The image of ${\cal V}_\Xi$ is a proper subspace of
${\cal V}_\Psi$
\begin{eqnarray}
S_0({\cal V}_\Xi) \subset
{\cal V}_\Psi.
\label{hier.5}
\end{eqnarray}
By construction 
\begin{eqnarray}
\Psi \in S_0({\cal V}_\Xi) 
\label{hier.6}
\end{eqnarray}
since there are no anomalies. It is convenient
to use a basis
\begin{eqnarray}
{\cal M}_ie_{ik} = S_0({\cal P}_k)
\label{hier.7}
\end{eqnarray}
where $k$ labels the chosen representatives of the
equivalence classes in ${\cal V}_\Xi$. Finally we have
\begin{eqnarray}
\Xi  = \sum_k \xi_k {\cal P}_k
\label{hier.8}
\end{eqnarray}
where $\xi_k$ are determined from
\begin{eqnarray}
\Psi  = \sum_{ki}{\cal M}_i e_{ik}\xi_k
\label{hier.9}
\end{eqnarray}
i.e.
\begin{eqnarray}
\psi_i=\sum_{k} e_{ik}\xi_k 
\label{hier.10}
\end{eqnarray}
In general the number of $\psi_i$ is higher than the number
of the unknowns $\xi_k$. The solution exists since the theory
is assumed to satisfy STI (no anomalies). Most of the consistency
conditions can be derived from the nilpotency of $S_0$
\begin{eqnarray}
S_0(\Psi) = 0.
\label{hier.11}
\end{eqnarray}
It should be noticed that $e_{ki}$ is a matrix fixed
by the model and by the choice of the basis 
$\left\{{\cal P}_k\right\}$. It can be evaluated
solely by using the BRS transformations given in Appendix 
\ref{sec:ST}. In particular it does
not depend on  the order of the perturbation expansion.
\par
The choice of the representatives and of the
linearly independent equations in (\ref{intr.8})
is performed according to the following strategy,
which aims to reduce the complexity of AR.
First, we look for a block or triangular structure of the matrix
$e_{ki}$ (hierarchy). Second, we reduce the  number of terms
coming from the lower perturbation expansion (see eq.
(\ref{intr.8})). Third, the choice of the linearly
independent equations is done by preferring the breaking
terms with lower number of external legs and higher 
derivatives in the external momenta. In this way the
number of graphs is reduced at the cost of some
derivatives on external momenta.
This strategy might look unnecessary in the present simple
model. However it will be useful in a more complicated
situation as, e.g., in the Standard Model.
\par
Two $A,B$ subspaces of ${\cal V}_\Xi$ are disjoint
   if
\begin{eqnarray}
S_0(A)\cap S_0(B) = \{0\}
\label{hier.12}
\end{eqnarray}
practically this means that the ST transforms of
$A,B$ do not shear any monomial ${\cal M}_i$. 
$A$ includes $B$ if
\begin{eqnarray}
S_0(B)\subset S_0(A) 
\label{hier.13}
\end{eqnarray}
These definitions are
the guide for the hierarchy structure of the
counterterms.  If they can be grouped into disjoint
sets then we have a block diagonalization of $e_{ki}$.
If we get an including structure then the matrix is
triangular. In both cases the task is consistently
reduced. Moreover we can use the ST invariants
in order to improve the structure of the matrices
$e_{ki}$ by choosing appropriate normalization
conditions. This is performed by exploiting the
invariance of eq. (\ref{intr.8}) under the transformation
\begin{eqnarray}
\Xi \to  \Xi + \sum_j v_j{\cal I}_j.
\label{hier.14}
\end{eqnarray}
The coefficients $v_j$ will be determined by excluding
some monomials ${\cal P}_k$ from the basis for $\Xi$.
%
%
%
\subsection{Ghost equation and invariant counterterms}
The proof of physical unitarity relay on the property of
$S$ of being nilpotent. In the present on-shell formalism
the ghost equation guarantees the above requirement
\begin{eqnarray}
\alpha\Box c + ev \Gamma_{J_2} = \Gamma_{\bar c}
\label{appa.11}
\end{eqnarray}
This requirement excludes a mass-term in $\Gamma^{(0)}$
of the form  
\begin{eqnarray}
M^2\left[{{A^2}\over 2}+{\bar c}c-
{1\over{2\alpha}}(\phi_1^2+\phi_2^2 )\right].
\label{appa.11p}
\end{eqnarray}
The present approach is equivalent to the Nakanishi-Lautrup
formulation of the gauge fixing\footnote{
The Nakanishi-Lautrup 
formulation requires a Lagrange multiplier $b$ coupled to the gauge 
fixing function ${\cal F}(A,\phi)$ (see the eq. ({\ref{40}})) and 
whose BRS transformation is simply given by $ s b= \bar{c}, s \bar{c} =0$. 
This provides an off-shell nilpotent BRS transformations 
avoiding the constraints (\ref{appa.11}) in order to guarantee the 
nilpotency of $S_0$.}.
The ghost  equation must be valid after the renormalization
procedure. For $n>1$ we have
\begin{eqnarray}
&& ev {{\rm I}\!\Gamma}_{J_2c}^{(n)} = {{\rm I}\!\Gamma}_{{\bar c}c}^{(n)}
\nonumber \\
&&
ev {{\rm I}\!\Gamma}_{J_2c\phi_1}^{(n)} = {{\rm I}\!\Gamma}_{{\bar c}c\phi_1}^{(n)}
\nonumber \\
&&
ev \Gamma_{J_2c\phi_1^2}^{(n)} = {{\rm I}\!\Gamma}_{{\bar c}c\phi_1^2}^{(n)}
\nonumber \\
&&
ev \Gamma_{J_2c\phi_2^2} ^{(n)}= {{\rm I}\!\Gamma}_{{\bar c}c\phi_2^2}^{(n)}
\nonumber \\
&&
ev \Gamma_{J_2cA_\mu^2}^{(n)}= {{\rm I}\!\Gamma}_{{\bar c}cA_\mu^2}^{(n)}
\label{appa.12}
\end{eqnarray}
These equations fix the counterterms
\begin{eqnarray}
\xi_{{\bar c}\Box c},
\xi_{{\bar c}c\phi_1^2},
\xi_{{\bar c}c\phi_2^2},
\xi_{{\bar c}cA_\mu^2}
\label{appa.13}
\end{eqnarray}
since they are related to superficially finite vertex functions.
The remaining counterterms
\begin{eqnarray}
\xi_{{\bar c}c},
\xi_{{\bar c}c\phi_1}
\label{appa.13a}
\end{eqnarray}
are related to counterterms involving external sources
\begin{eqnarray}
\xi_{J_2c},
\xi_{J_2 c\phi_1}.
\label{appa.13b}
\end{eqnarray}
Appendix \ref{sec:STI} list the linearly independent
ST invariants with charge conjugation +1. Any linear
combination of ST invariants
\begin{eqnarray}
\Xi \to  \Xi + \sum_{j=1,\dots,11} v_j{\cal I}_j.
\label{hier.14p}
\end{eqnarray}
can be added to the vertex functional. A straightforward
analysis shows that the ghost equation is preserved
provided
\begin{eqnarray}
v_7 = v_8 = 0
\label{appa.13c}
\end{eqnarray}
and moreover that, under such circumstances, the monomial
$\int d^4x{\bar c}\Box c $ is absent in the rest of the ST
invariants in eq. (\ref{hier.14p}).
\par
For further use we notice that the rest of the constants
$\{v_j\}$ can be determined by fixing the coefficients
of the following nine monomials
\begin{eqnarray}
\phi_1,\phi_2^2\phi_1,A^2\phi_1,
F_{\mu\nu}^2, i{\bar \psi}\gamma_5\psi \phi_2 ,
i{\bar \psi}\gamma_5\psi \phi_2 ,
J_2c ,J_2c\phi_1,J_1 c\phi_2
\label{appa.13d}
\end{eqnarray}
as can be seen from the matrix given in Appendix
(\ref{sec:STI}).
\subsection{Sector 0}
The counterterms containing external sources $J_i,\eta, {\bar\eta}$
are the right group to start with. 
\begin{eqnarray}
\xi_{J_2 c},\xi_{J_2 c\phi_1},
\xi_{J_1 c\phi_2},\xi_{{\bar\eta}c\psi},
\xi_{{\bar\psi}c\eta}.
\label{hier.15}
\end{eqnarray}
Their ST transforms (see Appendix \ref{sec:ST}) contain
the equations of motion and therefore one expects that they
belong to the subspace that includes (in the sense of eq. (\ref{hier.13}))
most of the subspaces of counterterms. 
Moreover in the recursive equation (\ref{hier.12}) the
counterterms which contain external sources 
are present in almost every terms.
Thus it is advantageous to set all possible loop corrections
to the BRS external sources  to zero by using the
freedom in the choice of the coefficients $\{v_j\}$ in eq.
(\ref{hier.14}). By using the ST invariants ${\cal I}_{9-11}$ given
in Appendix \ref{sec:STI}
we impose the normalization conditions ($n>0$)
\begin{eqnarray}
&&{{\rm I}\!\Gamma}_{J_2 c}^{(n)}(0)  =\xi^{(n)}_{J_2 c}  = 0
\nonumber\\ 
&&{{\rm I}\!\Gamma}_{J_2 c\phi_1}^{(n)}(0)  =\xi^{(n)}_{J_2 c\phi_1} = 0
\nonumber\\ 
&&{{\rm I}\!\Gamma}_{J_1 c\phi_2}^{(n)}(0)  =\xi^{(n)}_{J_1 c\phi_2} = 0.
\label{hier.16}
\end{eqnarray}
As a consequence of this choice eq. (\ref{appa.12})
now fixes the counterterms in eq. (\ref{appa.13a}), by using the
relation
\begin{eqnarray}
{{\rm I}\!\Gamma} = (1-t^{\delta_{\rm pc}})\Gamma + \Xi
\label{hier.16p}
\end{eqnarray}
one gets  $(n>0)$
\begin{eqnarray}
&&\xi^{(n)}_{{\bar c} c} ={{\rm I}\!\Gamma}_{{\bar c}c}^{(n)}(0)   = 0
\nonumber\\ 
&& 8
\xi^{(n)}_{{\bar c}\Box c} = \partial_{p_\mu}\partial_{p^\mu}
{{\rm I}\!\Gamma}_{{\bar c}c}^{(n)} (0) =
e v \partial_{p_\mu}\partial_{p^\mu}
\Gamma_{J_2c}^{(n)} (0)
\nonumber\\ 
&&\xi^{(n)}_{{\bar c} c\phi_1} =
{{\rm I}\!\Gamma}_{{\bar c} c\phi_1}^{(n)} (0) = 0
\nonumber\\ 
&&
\xi^{(n)}_{{\bar c} c\phi_1^2} = - e v \Gamma_{J_2c\phi_1^2}^{(n)}(0)
\nonumber\\ 
&&
\xi^{(n)}_{{\bar c} c\phi_2^2} = - e v \Gamma_{J_2c\phi_2^2}^{(n)}(0)
\nonumber\\ 
&&
\xi^{(n)}_{{\bar c} cA_\mu ^2}= - e v \Gamma_{J_2cA_\mu ^2}^{(n)}(0)
.
\label{hier.18}
\end{eqnarray}
Since the ghost equation fixes all counterterms involving the
ghost field, we drop the analysis of the ghost sector. The
ghost part of $\Xi$ is
\begin{eqnarray}
\Xi_{\rm GHOST} = 
\int d^4x \left[
\xi_{{\bar c}\Box c}{{\bar c}\Box c}
+\xi_{{\bar c} c\phi_1^2}{\bar c} c\phi_1^2
+\xi_{{\bar c} c\phi_2^2}{\bar c} c\phi_2^2
+\xi_{{\bar c} cA_\mu ^2}{\bar c} cA_\mu ^2
\right ].
\label{hier.19}
\end{eqnarray}
%
\subsection{Sector I}
The next sector is selected by the condition
\begin{eqnarray}
{\cal N}_\phi \leq 4,
{\cal N}_{A} = {\cal N}_\psi = {\cal N}_{\bar\psi} =  0
\label{hier.19p}
\end{eqnarray}
where ${\cal N}_\phi,{\cal N}_{A},{\cal N}_\psi$ 
and ${\cal N}_{\bar\psi}$ 
respectively count the number of $\phi, A, \psi, \bar{\psi}$.
The coefficients of of the monomial of this sector are
\begin{itemize}    
\item mass terms (3)\footnote{The number in brackets    
counts the number of counterterms of the corresponding sub-sector}:    
$\xi_{\phi_1}, \xi_{\phi_1\phi_1}, \xi_{\phi_2\phi_2}$     
\item trilinear self-interacting terms (2):    
$\xi_{\phi_1 \phi_2\phi_2}, \xi_{\phi_1\phi_1\phi_1}$    
\item quadrilinear interacting terms (3):     
$\xi_{\phi_1\phi_1\phi_2\phi_2},\xi_{\phi_1\phi_1\phi_1\phi_1 }, 
\xi_{\phi_2\phi_2\phi_2\phi_2}$   
\end{itemize}   
The sector can be further decomposed into two sub-sectors
with ${\cal N}_\phi \leq 2$ and ${\cal N}_\phi > 2$. These two
sub-sectors turn out to be disjoint if we put to zero the
coefficient $\xi_{\phi_1 \phi_2\phi_2}$ (see Appendix \ref{sec:ST}). 
This can be achieved by the ST invariant ${\cal I}_{2}$.
The contribution from the lower order of perturbation
are reduced if we put equal zero the coefficient
$\xi_{\phi_1}$ of the tadpole. The ST invariant
necessary to impose these conditions is ${\cal I}_{1}$.
Finally six coefficients have to be evaluated. A direct inspection
of the ST transforms of the corresponding monomial shows that
the breaking terms to be evaluated are six out of the following set
\begin{eqnarray}\psi^{I}_{i} = \left\{  \psi_{c \phi_2},  
\psi_{c \phi_2\phi_1},  \psi_{c \phi_2\phi_1^2}, 
\psi_{c \phi_2\phi_1^3}, \psi_{c \phi_2^3}, 
\psi_{c \phi_2^3\phi_1} \right\}.
\label{hier.20}
\end{eqnarray}
\par
With the above conventions it is straightforward, with the help
of the BRS transformations in Appendix \ref{sec:ST},
to construct the reduced matrix in eq. (\ref{hier.10})
\begin{eqnarray}
\left(
\begin{array}{lcccccc}
& \xi_{\phi_1^2} & 
\xi_{\phi_2^2} &  
\xi_{\phi_1^3}  &  
\xi_{\phi_1^2\phi_2^2} 
& \xi_{\phi_1^4 } &
\xi_{\phi_2^4} 
\vspace{.2cm} \\  
\psi_{c \phi_2}        & 0 &-2ev & 0 & 0 & 0 & 0  \\
\psi_{c \phi_2\phi_1}  &2e & -2e & 0 & 0 & 0 & 0 \\
\psi_{c \phi_2\phi_1^2}& 0 & 0 & 3e &- 2ev & 0 & 0  \\
\psi_{c \phi_2\phi_1^3}& 0 & 0 & 0 &- 2e & 4e & 0  \\
\psi_{c \phi_2^3}      & 0 & 0 & 0 & 0 & 0 & - 4ev \\
\psi_{c \phi_2^3\phi_1}& 0 & 0 & 0 & 2e & 0 & - 4e
\end{array}
\right)
\label{hier.21}
\end{eqnarray}
\subsection{Sector II}
This sector deals with the kinetic terms of the scalar fields and 
the corresponding 
terms coming form the covariant derivatives, 
that is the interaction terms of the 
scalar fields and the gauge fields. 
This sector also deals with the mass of the 
gauge bosons. This sector is selected by the condition
${\cal N}_\phi \leq 2, {\cal N}_\psi=0,{\cal N}_{\bar\psi}=0, 
{\cal N}_A+{\cal N}_\partial = 2$. $\xi^{II}$ are
\begin{itemize}    
\item mass term for gauge field (1):    
$\xi_{A_{\mu}^2}$     
\item kinetic terms for scalar fields (2):    
$ \xi_{\partial^\mu\phi_1\partial_\mu \phi_1},    
\xi_{\partial^\mu\phi_2\partial_\mu \phi_2}$  
\item mixing terms between scalar field and gauge field (1):     
$\xi_{\partial_\mu A^\mu \phi_2} $  
\item coupling scalar-gauge fields (2):    
$\xi_{A_{\mu}\partial^\mu \phi_2 \phi_1},    
\xi_{A_{\mu} \phi_2  \partial^\mu \phi_1}  $    
\item trilinear term (1):    
$\xi_{A_{\mu}^2\phi_1 }$     
\item quadrilinear terms (2):    
$\xi_{A_{\mu}^2 \phi_1^2}, \xi_{A_{\mu}^2 \phi_2^2}$ 
\end{itemize} 
The bilinear out-of-diagonal counterterm can be put to zero
\begin{eqnarray}
t^2 {{\rm I}\!\Gamma}_{A^{\mu} \phi_2}(0) =  
\Xi_{ A^{\mu} \phi_2}=0
\label{hier.21a}
\end{eqnarray}
by using the ST invariant ${\cal I}_3$.
Finally one has to evaluate eight coefficients
\begin{eqnarray}
\xi^{II} &\equiv &\Big\{
\xi_{A_{\mu}^2 },    
\xi_{\partial^\mu \phi_1\partial_\mu \phi_1},    
\xi_{\partial^\mu \phi_2\partial_\mu \phi_2} , 
\xi_{A_{\mu} \phi_2  \partial^\mu \phi_1},
\nonumber\\
&&
\xi_{A_{\mu}\partial^\mu \phi_2 \phi_1},    
\xi_{A_{\mu}^2\phi_1 },
\xi_{A_{\mu}^2 \phi_1^2}, \xi_{A_{\mu}^2 \phi_2^2}
\Big\}
\label{hier.25}
\end{eqnarray}
in terms of the following breaking terms
\begin{eqnarray}
\psi^{II}_i & = &  \left\{  
\psi_{\partial_\mu c\partial^\mu  \phi_2},  
\psi_{c \phi_2\Box \phi_1},  
\psi_{c\partial_\mu \phi_2\partial^\mu \phi_1},
\psi_{c\Box \phi_2 \phi_1},
\psi_{c\partial^\mu  A_{\mu}}, \right. \nonumber \\
&& \hspace{.2cm} \left. 
\psi_{c\partial^\mu  A_{\mu} \phi_1},
\psi_{c A^{\mu}\partial_\mu \phi_1},
\psi_{c\partial^\mu  A_{\mu} \phi_1^2},
\psi_{c A^{\mu}\partial_\mu \phi_1^2},
\psi_{c\partial^\mu A_{\mu} \phi_2^2}, 
\right. \nonumber \\ && \hspace{.2cm} \left. 
\psi_{c A^{\mu}\partial_\mu \phi_2^2}, 
\psi_{c A_{\mu} A^{\mu} \phi_1^2},
\psi_{c A_{\mu} A^{\mu} \phi_2^2},
\right\}
\label{hier.26}
\end{eqnarray}
The transformation matrix $e_{ik}$ (eq. (\ref{hier.10}))is 
\begin{eqnarray}
\left(
\begin{array}{lcccccccc}
&\xi_{A_{\mu}^2} &  
\xi_{\partial^\mu \phi_1\partial_\mu \phi_1}&
\xi_{\partial^\mu \phi_2\partial_\mu \phi_2}&
\xi_{A_{\mu} \phi_2  \partial^\mu \phi_1}&
\xi_{A_{\mu}\partial^\mu \phi_2 \phi_1}& 
\xi_{A_{\mu}^2\phi_1 }&
\xi_{A_{\mu}^2 \phi_1^2}& \xi_{A_{\mu}^2 \phi_2^2}
\vspace{.2cm}\\
\psi_{c\Box \phi_2} & 
0 & 0 & 2ev &  0 & 0 & 0 & 0 & 0 \\
\psi_{c \phi_2 \Box\phi_1} &
0& - 2e & 0 & 1 & 0 & 0 & 0 & 0  \\ 
\psi_{c\partial_\mu \phi_2\partial^\mu \phi_1} &
0 & 0 & 0 & 1 & 1 & 0 & 0   & 0  \\
\psi_{c\Box \phi_2 \phi_1} &
0 & 0 & 2e & 0 & 1 & 0 & 0   & 0  \\
\psi_{c\partial^\mu A_{\mu}} &
2 & 0 & 0 & 0 & 0 & 0 & 0 & 0  \\
\psi_{c \partial^\mu A_{\mu} \phi_1} &
0 & 0 & 0 & 0 & ev & 2 & 0  & 0  \\
\psi_{c A^{\mu} \partial_\mu \phi_1} &
0 & 0 & 0 & - ev  & ev & 2 & 0 & 0 \\
\psi_{c \partial^\mu  A_{\mu}\phi_1^2} &
0 & 0 & 0 & 0 & e & 0 & 2 & 0  \\
\psi_{c A^{\mu}\phi_1 \partial_\mu  \phi_1} &
0 & 0 & 0 & - e & e & 0 & 4  & 0 \\
\psi_{c  \partial^\mu A_{\mu} \phi_2^2} &
0 & 0 & 0 & - e & 0 & 0 & 0 & 2  \\ 
\psi_{c A^{\mu} \phi_2\partial_\mu  \phi_2} &
0 & 0 & 0 & - e & e & 0 & 0 & 4 \\ 
\psi_{c A_{\mu}^2 \phi_2} &
0 & 0 & 0 & 0 & 0 & e & 0 & -2ev  \\
\psi_{c A_{\mu}^2\phi_2\phi_1} &
0 & 0 & 0 & 0 & 0 & 0 & 2e & -2e
\end{array}
\right)
\nonumber \\
\label{hier.27}
\end{eqnarray}
%
\subsection{Sector III}
In the present model the kinetic terms for the 
gauge fields are trivial because of the abelianity of the gauge group. The 
eigenvalues of the counting operators are given by: 
\begin{eqnarray}
 {\cal N}_{\phi}=
 {\cal N}_{\psi} =  0, 
\left( {\cal N}_{A} + {\cal N}_{\partial} \right) 
 =  4 
\label{hier.28}
\end{eqnarray}
The sector contains
\begin{itemize}  
\item kinetic terms for gauge fields (2):    
$\xi_{\partial_\mu A^{\mu}\partial_\nu A^{\nu} }, 
\xi_{\partial_\nu A_{\mu}\partial^\nu A^{\mu} } $    
\item interacting terms (1):    
$\xi_{A_{\mu}^4 }$    
\end{itemize}    
The corresponding breaking-terms are given by:
\begin{eqnarray}
\psi^{III}_{i} & = &  \left\{  
\psi_{c \Box\partial_\mu A^{\nu}}, 
\psi_{c\partial^\mu A_{\mu} A_{\nu}^2},
\psi_{c\partial^\nu A_{\mu} A_{\nu}A^\mu}
\right\}
\label{hier.29} 
\end{eqnarray}
The ST invariant  ${\cal I}_4$ can be used in order to put equal zero
the counterterm corresponding to the  transverse part
\begin{eqnarray}
\xi_{F_{\mu\nu}^2} = 0.
\label{hier.30} 
\end{eqnarray}
By looking at eq. (\ref{appb.42}) we have the following relations
\begin{eqnarray}
&&
\psi_{c \Box\partial_\mu A^{\nu}} = 
- 2\xi_{\partial_\mu A^{\mu}\partial_\nu A^{\nu} }
\nonumber\\
&&
\psi_{c\partial^\mu A_{\mu} A_{\nu}^2} =
4 \xi_{A_{\mu}^4 } - \xi_{{\bar c}cA_\mu^2}
\nonumber\\
&&
\psi_{c\partial^\nu A_{\mu} A_{\nu}A^\mu} = 
8 \xi_{A_{\mu}^4 }
\label{hier.31} 
\end{eqnarray}
where the coefficient $\xi_{{\bar c}cA_\mu^2}$ is known from the
ghost equation (\ref{hier.18}).
%
\subsection{Sector IV}
This sector contains the Green's functions with fermion 
fields, and it can be further divided into the sector of mass terms of fermion 
fields and their coupling with the scalar fields and the sector of the 
kinetic terms and the interaction with the gauge fields. The present sector 
is completely decoupled for the previous sectors, 
it is specified by the following  eigenvalues:
\begin{eqnarray}
{\cal N}_{\phi} \leq 1 ,
{\cal N}_{\psi}  =  2,
{\cal N}_{A} + {\cal N}_{\partial}  \leq 1 
\label{hier.32} 
\end{eqnarray}
and the counterterms are 
\begin{itemize}
\item mass term (1): $\xi_{{\bar\psi} \psi}$ 
\item Yukawa term (2): $\xi_{{\bar\psi} \psi \phi_1}, 
\xi_{i{\bar\psi}\gamma_5 \psi \phi_2}$ 
\item kinetic term  and interaction  with the gauge field (2):
$\xi_{i{\bar\psi}{\not \partial }  \psi}, \xi_{{\bar\psi} \not \! A\psi }$ 
\end{itemize} 
The breaking-terms are given by:
\begin{eqnarray}
\psi^{IV}_{i} & = &  \left\{  
\psi_{c {\bar\psi}\gamma_5 \psi}, \psi_{c \phi_1 {\bar \psi}\gamma_5 \psi}, 
\psi_{c \phi_2 {\bar \psi} \psi},  
\psi_{c {\bar \psi}\gamma_5{\not \partial} \psi}, 
\psi_{c\partial^\mu {\bar \psi}\gamma_\mu\gamma_5 \psi},
\right\}
\label{hier.33}
\end{eqnarray}
There are two  invariants (${\cal I}_5$ and ${\cal I}_6$),
pertinent to this sector.
They are used to impose the following normalization conditions
\begin{eqnarray}
&&
\xi_{{\bar\psi} \not \! A\psi } = 0
\nonumber\\
&&
\xi_{i{\bar\psi}\gamma_5 \psi \phi_2} = 0.
\label{hier.34}
\end{eqnarray}
The matrix $e_{ik}$ which express the functional $\psi^{(IV)}$ in terms
of $\xi^{(IV)}$ is given by
\begin{eqnarray}
\left(
\begin{array}{lccc}
&\xi_{{\bar\psi} \psi} & 
\xi_{{\bar\psi} \psi \phi_1} &  
\xi_{i{\bar\psi}{\not \partial }  \psi}
\vspace{.2cm} \\  
\psi_{c {\bar\psi}\gamma_5 \psi}          & -1 & 0 & 0 \\
\psi_{c \phi_1 {\bar \psi}\gamma_5 \psi}  & 0 & -1 & 0 \\
\psi_{c \phi_2 {\bar \psi} \psi}          & 0 & 1 & 0 \\
\psi_{c {\bar \psi}\gamma_5{\not \partial} \psi}& 0 & 0 & - {1\over 2}  \\
\psi_{c\partial^\mu {\bar \psi}\gamma_\mu\gamma_5 \psi}& 0 & 0 &
- {1\over 2}
\end{array}
\right)
\label{hier.35}
\end{eqnarray}
\subsection{Summary of the normalization conditions}
For $n>0$ we have imposed the normalization conditions
\begin{eqnarray}
\begin{array}{llllll}
\xi^{(n)}_{\phi_1} = 0
&
\xi^{(n)}_{\phi_2^2\phi_1} = 0
&
\xi^{(n)}_{\partial^\mu A_\mu\phi_2} = 0
&
\xi^{(n)}_{F^2_{\mu\nu}} = 0
&
\xi^{(n)}_{i{\bar\psi}\gamma_5 \psi \phi_2} = 0
&
\xi^{(n)}_{{\bar\psi} \not \! A\gamma_5\psi } = 0
\\
\xi^{(n)}_{J_2 c}  = 0
&
\xi^{(n)}_{J_2 c\phi_1} = 0
&
\xi^{(n)}_{J_1 c\phi_2} = 0.
& & &
\end{array}
\label{hier.36}
\end{eqnarray}
The evaluation of physical S-matrix elements requires the evaluations
of the eigenvalues and of the eigenvectors of the two-points
vertex functions. 
The physical amplitudes are then obtained from the connected
and truncated Feynman amplitudes evaluated on the physical states
obtained from the diagonalization procedure (LSZ reduction formalism).
Thus on-shell normalization can be by-passed. The coupling constants
and masses 
in $\Gamma^{(0)}$ are dummy parameters which can be
obtained from a sufficient number of physical processes.

\section{ST breaking terms}
\label{sec:bterm}
In the strategy outlined before the counterterm functional $\Xi$
is obtained by solving a set of linear equations
(\ref{hier.10}). The restoration of ST invariance 
consists in the evaluation
of a certain number of (finite) vertex functions (the functional
$\Psi$). This fact
puts in clear evidence that it is the finite part of the
perturbative expansion that fixes the counterterms in the action.
\par
In this section we discuss some aspects of this procedure.
The first step consists in the evaluation of the functional
derivatives of $\Psi$. It is of some help to remember that,
in absence of anomalies,
$\Psi$ is the image through $S_0$ of non-invariant counterterms ($\Xi$).
Therefore it has FP-charge equal $+1$, $C=0$ and dimension less
or equal five. The next step is to find the coefficients $\psi_i$
in the expansion in terms of Lorentz scalar monomials
\begin{eqnarray}
\Psi = \sum_i \psi_i {\cal M}_i
\label{break.0}
\end{eqnarray}
\par
 Let us write explicitly,
for $n>0$, the operator $S_0$, where we impose the ghost equation
of motion given in eq. (\ref{appa.11})
i.e.
\begin{eqnarray}
ev \Gamma_{J_2}^{(n)} = \Gamma_{\bar c}^{(n)}
\qquad {\rm for } \quad n>0
\label{break.00}
\end{eqnarray}
\begin{eqnarray}
{\hat S}_0(\Gamma^{(n)})&\equiv&
\int d^4x \left\{ \partial_\mu c\Gamma^{(n)}_{A_\mu}
-ec\phi_2\Gamma^{(n)}_{\phi_1}
+ ec(\phi_1+v)\Gamma^{(n)}_{\phi_2}
+ i {e\over 2} c{\bar\psi} \gamma_5 \Gamma^{(n)}_{\bar \psi}
+ i{e\over 2} c\gamma_5\psi \Gamma^{(n)}_{\psi}
\right. \nonumber \\
&+& \left. 
\Gamma^{(0)}_{\phi_1}\Gamma^{(n)}_{J_1}
+ \left[\Gamma^{(0)}_{\phi_2}
+ev(\partial^\mu A_\mu + {{ev}\over\alpha}\phi_2)\right]\Gamma^{(n)}_{J_2}
- \Gamma^{(0)}_{\psi}\Gamma^{(n)}_{\bar \eta}
+ \Gamma^{(0)}_{\bar \psi}\Gamma^{(n)}_{\eta} \right\}
\nonumber \\
\label{break.1}
\end{eqnarray}
By imposing the condition (\ref{break.00}) the breaking
term $\Psi$ changes. We denote this change with the
notation $\Psi\to {\hat\Psi}$. 
\par
In the linearized form ($S_0$) one of the factors in each
monomial contains the vertex function at zero loop $\Gamma^{(0)}$.
All these facts have some interesting consequences
\begin{enumerate}
\item The functional derivatives of $\Psi$ relevant for the
evaluation of the counterterm can be read directly from the
BRS transforms of all action-like terms (see Appendix
\ref{sec:ST}).
\item Let $\delta$ be the total dimension of the fields we use for
the functional derivative of $\Psi$. Then the order of the Taylor 
operator  $\delta_{\rm D}$ is (see eq. (\ref{intr.3}))
\begin{eqnarray}
\delta_{\rm D} = 5-\delta
\label{break.1a}
\end{eqnarray}
\item Let us consider a generic term of $S_0$ for instance
$\Gamma^{(0)}_{J_i}\Gamma^{(n)}_{\phi_i}$ or 
$\Gamma^{(n)}_{J_i}\Gamma^{(0)}_{\phi_i}$. If 
$\Gamma^{(0)}$ does not contain any dimensional parameter,
then 
\begin{eqnarray}
t^{\delta_{\rm D}} \Gamma^{(0)}_{J_i}\Gamma^{(n)}_{\phi_i} = 
\Gamma^{(0)}_{J_i}t^{\delta_{\rm pc}} \Gamma^{(n)}_{\phi_i}.
\label{break.1b}
\end{eqnarray}
In the above equation we use a rather short-hand writing and 
to be more explicit we give an
example: by taking the functional derivative of the 
$\Gamma^{(0)}_{J_1}\Gamma^{(n)}_{\phi_1}$ term with respect to
$c\phi_2$, the $\delta_{\rm D}$ is equal to $3$ and we get 
\begin{eqnarray}
t^3 \Gamma^{(0)}_{J_1c\phi_2}\Gamma^{(n)}_{\phi_1} = -e
t^3 \Gamma^{(n)}_{\phi_1}.
\label{break.1c}
\end{eqnarray}
where 
$\delta_{\rm pc}=3$ is the superficial degree of divergence
of $\Gamma^{(n)}_{\phi_1}$.
\item The above point implies that $\Psi$ (eq. (\ref{intr.7}))
gets contributions only from those terms of $\Gamma^{(0)}$
which carry a dimensioned parameter ($v$ and masses).
\end{enumerate}
The functional derivatives of $\Psi$ are performed in Appendix
\ref{sec:psi}. It should be noticed that, due to our choice
of normalization conditions, few other counterterms turn out
to be zero at every order. 
\begin{eqnarray}
\int d^4 x \phi_1^2 \quad
\int d^4 x \phi_2^2 \quad
\int d^4 x (\partial_\mu\phi_2)^2 \quad
\int d^4 x A_\mu^2.
\label{break.1d}
\end{eqnarray}
This is due to the combined effects 
of our choice
of normalization conditions and of the 
zero momentum subtraction procedure.
Moreover the contribution to STI
from the lower order amplitudes appear only in few functional
derivatives of $\Psi$. One can describe this fact by saying
that the set of STI becomes {\sl almost} linear in $\Gamma$.
\section{Solution for counterterms}
\label{sec:sol}
The relations obtained in Appendix \ref{sec:psi} can be expanded 
in terms of Lorentz invariant amplitudes. Thus one can express 
the invariant amplitude for $\Xi$ in terms of the invariant 
amplitude for $\Psi$. This amounts to solve the linear algebra problem 
given in equation (\ref{hier.10}) where the matrices are given in (\ref{hier.21}), 
(\ref{hier.27}), (\ref{hier.21}), (\ref{hier.31}) and (\ref{hier.35}).
We remind our notations where the small letters $\xi$, $\gamma$ and $\psi$ denote 
the coefficients of the Lorentz invariant monomials 
respectively of  
of $\Xi$ (counterterms), $\Gamma$ and $\Psi$ (breaking terms)
indicated by the subscript. 
The order of perturbation theory is not shown and it is understood to
be $n$, unless explicitly exhibited.
\par
The ghost equation  (\ref{appa.11}) fixes the kinetic counterterms of the ghost
( see eq. (\ref{hier.18})) 
\begin{eqnarray}
\xi_{{\bar c}\Box c} = ev\gamma_{ \Box J_2c}
\label{cou.1}
\end{eqnarray}
\subsection{Counterterms of sector I}
In this sector we have the same number of equations and unknowns. 
The solution is (including the normalization conditions) 
\begin{eqnarray}
&& \xi_{\phi_1} =0 \quad \xi_{\phi_2^2} =0 \quad \xi_{\phi_1^2} =0  \quad 
\xi_{\phi_2^2\phi_1} =0  \quad \xi_{\phi_2^4}= 0 
\nonumber\\&&
\xi_{\phi_1^3} = {1\over{3e}} \left\{
-m_1^2 \gamma_{ J_2 c\phi_2^2}
+4ev^2\gamma_{\phi_2^4 \phi_1}
-m_1^2 v\gamma_{J_1 c\phi_2^3}
-{3\over 2}m_1^2\gamma_{J_1 c\phi_2\phi_1}
\right\}
\nonumber\\&&
\xi_{\phi_2^2\phi_1^2}= {1\over{2e}} \left\{
-2\lambda v \gamma_{ J_2 c\phi_2^2}
-v\lambda\gamma_{J_1 c\phi_2\phi_1}
+4ev\gamma_{\phi_2^4 \phi_1}
-m_1^2 \gamma_{J_1 c\phi_2^3}
\right\} 
\nonumber\\&&
\xi_{\phi_1^4}
={1\over{4e}}\Big\{
-2v\lambda\gamma_{J_2 c\phi_1^2}
-4v\lambda\gamma_{J_1 c\phi_2\phi_1}
-m_1^2\gamma_{J_1 c\phi_2\phi_1^2}
+2ev \gamma_{\phi_2^2 \phi_1^3}
\nonumber\\&& \quad
-2\lambda v \gamma_{ J_2 c\phi_2^2}
+4ev\gamma_{\phi_2^4 \phi_1}
-m_1^2 \gamma_{J_1 c\phi_2^3}
+3\sum_{j=1}^{n-1}\gamma^{(j)}_{J_1c\phi_2\phi_1}\xi^{(n-j)}_{\phi_1^3}
\Big\}
\nonumber\\&&
\label{expl.1}
\end{eqnarray}
\subsection{Counterterms of sector II}
In this sector the problem is over-determined. 
We use the first six and the last two rows of the matrix (\ref{hier.27}) .
The solution is (including the normalization conditions) 
\begin{eqnarray}
&& \xi_{A^2}= 0 \qquad \xi_{A^\mu\partial_\mu\phi_2}= 0  \qquad 
\xi_{\partial_\mu\phi_2\partial^\mu\phi_2}
= 0 
\nonumber\\&&
\xi_{A^\mu\phi_1\partial_\mu\phi_2}=
-m_1^2\gamma_{ J_1c\Box\phi_2}
-2ev \gamma_{\partial_\mu\phi_2\partial^\mu\phi_2\phi_1}
-2\lambda v \gamma_{\Box J_2c} 
\nonumber\\&&
\xi_{A^\mu\partial_\mu\phi_1\phi_2}=
m_1^2\gamma_{ J_1c\Box\phi_2}
-m_1^2\gamma_{\partial_\mu J_1c\partial^\mu\phi_2}
-2\lambda v \gamma_{\Box J_2c}
\nonumber\\&&
\xi_{\partial_\mu\phi_1\partial^\mu\phi_1}={1\over{2e}}
\Big\{
m_1^2\gamma_{\Box J_1c\phi_2}
+ev \gamma_{\partial_\mu\phi_2\phi_2\partial^\mu\phi_1}
-m_1^2\gamma_{\partial_\mu J_1c\partial^\mu\phi_2}
+m_1^2\gamma_{ J_1c\Box\phi_2}
\Big\}
\nonumber\\&&
\xi_{A^2\phi_1} = 
-{1\over{2}}m_1^2 \gamma_{J_1c\partial_\mu  A^\mu}
+{1\over{2}}m_1^2ev\gamma_{ J_1c\Box\phi_2}
+(ev)^2 \gamma_{\partial_\mu\phi_2\partial^\mu\phi_2\phi_1}
+{1\over{2}}m_1^2e \gamma_{\Box J_2c}
\nonumber\\&&
\xi_{A^2\phi_2^2} = {1\over{2v}} \xi_{A^2\phi_1}
\nonumber\\&&
\xi_{A^2\phi_1^2}
={1\over{2e}} \Big\{
-m_1^2 \gamma_{ J_1 c\phi_2 A^2}
-2v\lambda\gamma_{J_2 cA^2}
+e^2v\gamma_{J_1 c\phi_2\phi_1}
\nonumber\\&&
-\lambda ev  \gamma_{J_1c\partial_\mu  A^\mu}
+ \lambda (ev)^2 \gamma_{ J_1c\Box\phi_2}
+e^3 v\gamma_{\partial_\mu\phi_2\partial^\mu\phi_2\phi_1}
+\lambda e^2 v\gamma_{\Box J_2c}
+\sum_{j=1}^{n-1}\gamma^{(j)}_{J_1c\phi_2\phi_1}\xi^{(n-j)}_{\phi_1 A^2}
\Big\}
\label{expl.9}
\end{eqnarray}
The rest of the equations provided by the matrix (\ref{hier.27}) gives 
consistency conditions. However not all of them are linear independent, 
in fact one can easily check that the linear 
combination implied by 
\begin{eqnarray}
-2\psi_{c A^2\phi_2}
+e\psi_{c A^\mu\partial_\mu \phi_1}
-ev\psi_{c A^\mu\partial_\mu \phi_2\phi_2}
\label{pp.12}
\end{eqnarray}
gives an identity. It should be reminded that this peculiar property 
is a consequence of our normalization conditions. Then the consistency conditions 
are 
\begin{enumerate}
\item
\item
\begin{eqnarray}
&&
ev [\gamma_{\partial_\mu J_1c\partial^\mu\phi_2}
-\gamma_{ J_1c\Box\phi_2} ]
+ e \gamma_{\Box J_2c}
+\gamma_{\partial_\mu J_1cA^\mu}
-\gamma_{J_1c\partial_\mu  A^\mu} 
=0
\label{newexpl.16}
\end{eqnarray}
\item
\begin{eqnarray}
&&
6\lambda v(\gamma_{J_1c\partial_\mu  A^\mu}
-\gamma_{\partial_\mu J_1 c A^\mu})
-m_1^2 \gamma_{J_1 c A^\mu \partial_\mu\phi_1}
-m_1^2 \gamma_{\partial_\mu J_1 c A^\mu \phi_1}
+2m_1^2 \gamma_{J_1c\partial_\mu  A^\mu\phi_1}
-m_1^2e\gamma_{\partial_\mu J_1c\partial^\mu\phi_2}
\nonumber\\&&
-2e^2v \gamma_{\partial_\mu\phi_2\partial^\mu\phi_2\phi_1}
-4\lambda ev \gamma_{\Box J_2c}
- 2ev \gamma_{J_2 c \phi_1^2}
+2ev (\gamma_{\phi_2\partial_\mu\phi_1^2  A^\mu}
-\gamma_{\phi_2\phi_1^2 \partial_\mu A^\mu})=0
\label{newcc.6}
\end{eqnarray}
\item
\begin{eqnarray}
&& +\lambda v^2 e^2 \gamma_{ J_1c\Box\phi_2}
X+e^3v \gamma_{\partial_\mu\phi_2\partial^\mu\phi_2\phi_1}
+\lambda ve^2 \gamma_{\Box J_2c}
-2\lambda ve \gamma_{J_1c\partial_\mu  A^\mu}
+e^2v \gamma_{\phi_2\phi_1^2 \partial_\mu A^\mu}
\nonumber\\&&
+2\lambda v^2 \gamma_{ J_1 c\phi_2 A^2}
+2v\lambda\gamma_{J_2 cA^2}
-e^2v\gamma_{J_1 c\phi_2\phi_1}
-2\lambda v^2e\gamma_{J_1c\partial_\mu  A^\mu\phi_1}
+e^2v \gamma_{J_2 c \phi_1^2} 
\nonumber\\&&
-\sum_{j=1}^{n-1}\gamma^{(j)}_{J_1c\phi_2\phi_1}\xi^{(n-j)}_{\phi_1 A^2}
+3e\sum_{j=1}^{n-1} \gamma^{(j)}_{J_1c\partial^\mu  A_\mu}\xi^{(n-j)}_{\phi_1^3}
= 0
\label{newcons.40}
\end{eqnarray}
%
\begin{eqnarray}
&&
+3\gamma_{\phi_2^3\partial_\mu A^\mu}
+\lambda v \gamma_{ J_1c\Box\phi_2}
-2\lambda v \gamma_{\partial_\mu J_1c\partial^\mu\phi_2}
-3\lambda   \gamma_{\Box J_2c}
+ \gamma_{ J_2 c\phi_2^2}
-e \gamma_{\partial_\mu\phi_2\partial^\mu\phi_2\phi_1}
=0
\label{newcons.42}
\end{eqnarray}
\end{enumerate}
\subsection{Counterterms of sector III}
The counterterms of the sector III, together with the normalization 
condition, are 
\begin{eqnarray}
&& \xi_{F_{\mu\nu}^2} = 0 
\nonumber\\&&
\xi_{\partial_\mu A_\nu \partial^\nu A^\mu}={1\over 2}\Big\{
ev\gamma_{A_\mu\partial^\mu \Box\phi_2 } -ev\gamma_{ \Box J_2c}
\Big\}
\nonumber\\&&
\xi_{ A^4} ={1\over 4}\Big\{
ev \gamma_{\phi_2 \partial^\mu A_\mu A^2}
+e^2v\gamma_{\partial_\mu  J_1 cA^\mu}
+\sum_{j=1}^{n-1}\gamma^{(j)}_{\partial^\mu J_1c  A_\mu}\xi^{(n-j)}_{\phi_1 A^2}
\Big\}
\label{expl.17a}
\end{eqnarray}
In this sector there is one consistency condition 
\begin{eqnarray}
&&
\gamma_{ \phi_2  A_\mu\partial^\mu A^2}
+e\gamma_{ J_1 c\partial_\mu A^\mu}
+\gamma_{ J_2 cA^2}
+\sum_{j=1}^{n-1}\gamma^{(j)}_{J_1c\partial^\mu A_mu}
\xi^{(n-j)}_{\phi_1 A^2}
\nonumber\\&&
= 
\gamma_{\phi_2 \partial^\mu A_\mu A^2}
+e\gamma_{\partial_\mu  J_1 cA^\mu}
+\sum_{j=1}^{n-1}\gamma^{(j)}_{\partial^\mu J_1c A_\mu}
\xi^{(n-j)}_{\phi_1 A^2}.
\label{newcons.50}
\end{eqnarray}
\par
It is remarkable that contribution from the lower order 
terms appear only in three counterterms (see the eqs. 
(\ref{expl.1}), (\ref{expl.9}) and (\ref{expl.17a})). 
\subsection{Counterterms of the fermion sector}
The analysis of the breaking terms in terms of the Lorentz invariant 
amplitude performed in the Appendix \ref{sec:psi} reveals 
that the fermion source counterterms are non vanishing.
The counterterms of this sector are 
\begin{eqnarray}
&&
\xi_{{i\over 2} ({\bar\eta}\gamma_5\psi c
+c{\bar\psi}\gamma_5\eta)}= - 
{1\over{2G}}\Big\{
2ev \gamma_{\phi_2 \phi_2{\bar\psi} \psi}
-2Gv \gamma_{c\phi_2{\bar \eta}\psi }
\nonumber\\&&
+ev \gamma_{i\phi_2 \phi_1{\bar\psi}\gamma_5  \psi}
-2Gv \gamma_{ic\phi_1{\bar\eta }\gamma_5\psi }
-m_1^2\gamma_{iJ_1c{\bar\psi}\gamma_5  \psi} 
\nonumber\\&&
-\sum_{j=1,n-1}\Big[2 \xi_{{\bar \psi} \psi}^{(j)} \Big(
\gamma_{c\phi_2{\bar\psi }\eta }^{(n-j)}
+\gamma_{ic\phi_1{\bar\psi}\gamma_5\eta }^{(n-j)}\Big)
+\xi_{\phi_1{\bar\psi}\psi}^{(j)}
\xi_{{i\over 2} ({\bar\eta}\gamma_5\psi c
+c{\bar\psi}\gamma_5\eta)}^{(n-j)}
\Big]
\Big\}
\label{fermi.30p}
\end{eqnarray}
The other counterterms are
\begin{eqnarray}
&&
\xi_{\phi_1{\bar \psi} \psi}= 
{1\over{2e}}\Big\{
2ev \gamma_{\phi_2 \phi_2{\bar\psi} \psi}
-2Gv \gamma_{c\phi_2{\bar \eta}\psi }
\nonumber\\&&
-ev \gamma_{i\phi_2 \phi_1{\bar\psi}\gamma_5  \psi}
+2Gv \gamma_{ic\phi_1{\bar\eta }\gamma_5\psi }
+m_1^2\gamma_{iJ_1c{\bar\psi}\gamma_5  \psi} 
\nonumber\\&&
-\sum_{j=1,n-1}\Big[2 \xi_{{\bar \psi} \psi}^{(j)} \Big(
-\gamma_{c\phi_2{\bar\psi }\eta }^{(n-j)}
+\gamma_{ic\phi_1{\bar\psi}\gamma_5\eta }^{(n-j)}\Big)
+\xi_{\phi_1{\bar\psi}\psi}^{(j)}
\xi_{{i\over 2} ({\bar\eta}\gamma_5\psi c
+c{\bar\psi}\gamma_5\eta)}^{(n-j)}
\Big]
\Big\}
\label{fermi.31p}
\end{eqnarray}
\begin{eqnarray}
\xi_{{\bar \psi} \psi} = - 
{{1}\over e}\Big[
Gv\xi_{{i\over 2} ({\bar\eta}\gamma_5\psi c
+c{\bar\psi}\gamma_5\eta)}
-\sum_{j=1,n-1}  
\xi_{{\bar\psi}\psi}^{(j)}
\xi_{{i\over 2} ({\bar\eta}\gamma_5\psi c
+c{\bar\psi}\gamma_5\eta)}^{(n-j)}
\Big]
\label{fermi.32}
\end{eqnarray}
and
\begin{eqnarray}
&&
\xi_{i{\bar\psi}\gamma_\mu\partial^\mu\psi}=
{2\over e}\Big\{
- {1\over 2}\xi_{{i\over 2} ({\bar\eta}\gamma_5\psi c
+c{\bar\psi}\gamma_5\eta)}
\nonumber\\&&
+Gv\gamma_{{\bar \eta}c\gamma_\mu\gamma_5\partial^\mu\psi}
+Gv \gamma_{{\bar\psi} c\gamma_\mu  \gamma_5\partial^\mu\eta}
-ev \gamma_{\phi_2 {\bar\psi} \gamma_\mu  \gamma_5\partial^\mu\psi}
\nonumber\\&&
-\sum_{j=1,n-1}\Big[
\xi_{{\bar\psi}\psi}^{(j)}\Big(
\gamma_{c{\bar \psi}\gamma_\mu\gamma_5\partial^\mu\eta }^{(n-j)}
-\gamma_{c{\bar\eta}
\gamma_\mu\gamma_5\partial^\mu \psi}^{(n-j)})
\nonumber\\&&
+{1\over 2}  \xi_{i{\bar\psi}\gamma_\mu\partial^\mu\psi}^{(j)}
\xi_{{i\over 2} ({\bar\eta}\gamma_5\psi c
+c{\bar\psi}\gamma_5\eta)}^{(n-j)}
\Big]
\Big\}
\label{fermi.23}
\end{eqnarray}
Since we require hermiticity and charge conjugation invariance, 
there is one consistency condition left, given by the equation 
\begin{eqnarray}
&&
0 = - ev \gamma_{\phi_2 \gamma^{\mu} A_\mu {\bar \psi} \psi} + 
G\, v \gamma_{c {\bar \eta } \gamma^{\mu} \gamma^{5} \psi A_\mu} - 
G\, v \gamma_{c {\bar \psi } \gamma^{\mu} \gamma^{5} \eta A_\mu} + 
\nonumber\\&& +
\sum_{j=1,n-1}\Big( \xi_{{\bar \psi} \psi}^{(j)}  
\gamma^{(n-j)}_{c {\bar \eta } \gamma^{\mu} \gamma^{5} \psi A_\mu} - 
\xi_{{\bar \psi} \psi}^{(j)} \gamma^{(n-j)}_{c {\bar \psi } \gamma^{\mu} \gamma^{5} \eta A_\mu}
\Big)
\label{fermi.33}
\end{eqnarray}

\section{Conclusions}
The absence of anomalies in the Higgs-Kibble model 
allows the explicit construction of counterterms which 
re-establish the Slavnov-Taylor invariance of the model. Therefore 
any regularization procedure which preserves the Lorentz covariance and 
the relevant discrete symmetries can be corrected by finite counterterms.  
In the present work we 
give explicitly  the counterterms in terms 
of a set finite vertex functions. Our strategy relies on two 
essential ingredients. One is the possibility to perform 
subtraction at zero momentum. The second consists 
in the use of the normalization conditions which simplify the 
construction of explicit solutions. Quite a few counterterms 
turn out to be zero and moreover the contribution of the 
lower terms in the perturbative expansion is highly reduced. 
Although the solution look cumbersome we believe that it
makes possible 
the automatic evaluation of the counterterms. 
%
\section{Acknowledgments} 
We would like to thank C. Becchi from drawing our attention
on the problem of the actual construction of Algebraic
Renormalization and
D.Maison and W.Zimmermann for useful discussions. 
One of us (R.F.) wishes to thank Max-Planck-Institut f\"ur Physik 
Werner-Heisenberg-Institut for its hospitality. Financial support by 
MURST is acknowledged. 
\appendix
\section{Classical action and BRS}
\label{sec:class}
\subsection{Feynman rules}
The Lagrangian density is
\begin{eqnarray}
&&
{\cal L}=  - {1\over 4}F_{\mu\nu}^2
- {\alpha\over 2}(\partial A)^2 + |D_\mu\phi|^2
- \lambda (|\phi|^2-{{v^2}\over2})^2
\nonumber\\
&&
+ {\bar\psi}i\not \!\!{\cal D}\psi 
+{{G}\over{\sqrt{2}}}{\bar\psi}(1-\gamma_5)\psi\phi
+{{G}\over{\sqrt{2}}}{\bar\psi}(1+\gamma_5)\psi\phi^*
\label{35}
\end{eqnarray}
where
\begin{eqnarray}
&&D_\mu = \partial_\mu - i e A_\mu
\nonumber\\
&&
{\cal D} - \partial_\mu - i {e\over 2} \gamma_5 A_\mu
\label{36}
\end{eqnarray}
BRS transformations
\begin{eqnarray}
&&
\delta A_\mu = \partial_\mu c
\nonumber\\
&&
\delta \phi = i e c \phi
\nonumber\\
&&
\delta \phi^* = -i e c \phi^*
\nonumber\\
&&
\delta \psi = -i {e\over 2} \gamma_5 \psi c
\nonumber\\
&&
\delta {\bar\psi} = i {e\over 2} c{\bar\psi} \gamma_5 
\label{37}
\end{eqnarray}
Now we consider the spontaneous symmetry breaking
\begin{eqnarray}
\phi = {{\phi_1 + v + i \phi_2}\over \sqrt{2}}
\label{38}
\end{eqnarray}
The bilinear parts give a out-of-diagonal term
\begin{eqnarray}
ev \phi_2 \partial A
\label{39}
\end{eqnarray}
thus we need a gauge fixing ('t Hooft)
\begin{eqnarray}
-{\alpha\over 2}\left(\partial A +
{{ev}\over{\alpha}}\phi_2\right)^2 
\label{39a}
\end{eqnarray}
Thus we complete the BRS
\begin{eqnarray}
&&
\delta \phi_1 = - e c \phi_2
\nonumber\\
&&
\delta \phi_2 =  e c (\phi_1+v)
\nonumber\\
&&
\delta {\bar c} = {\cal F} = \partial A +
{{ev}\over{\alpha}}\phi_2
\label{40}
\end{eqnarray}
Then the gauge fixing term is 
\begin{eqnarray}
&&
\Gamma^{(0)}_{\rm GF} = \int d^4x \Big [
-{\alpha\over 2}{\cal F}^2 +\alpha {\bar c}
\delta {\cal F}\Big ]
\nonumber\\
&&
=\int d^4x \Big [
-{\alpha\over 2}{\cal F}^2 +\alpha {\bar c}
\Box c + (ev)^2 {\bar c}c + e^2 v {\bar c}c\phi_1
\Big ]
\label{41}
\end{eqnarray}
and the zero-loop action is 
\begin{eqnarray}
&&
\Gamma^{(0)} = \int d^4x \Big [- {1\over 4}F_{\mu\nu}^2
+{{e^2v^2}\over 2}A_\mu^2
\nonumber\\
&&
-{\alpha\over 2}\partial A^2 + \alpha {\bar c}\Box c 
+ (ev)^2 {\bar c}c + e^2 v {\bar c}c\phi_1
\nonumber\\
&&
+{1\over 2}(\partial_\mu\phi_1^2 + \partial_\mu\phi_2^2)
-\lambda v^2 \phi_1^2 - {{(ev)^2}\over {2\alpha}}\phi_2^2
\nonumber\\
&&
+eA_\mu (\phi_2\partial^\mu \phi_1-\partial^\mu\phi_2\phi_1)
+ e^2v\phi_1 A^2 +{{e^2}\over 2}(\phi_1^2+\phi_2^2) A^2 
\nonumber\\
&&
- \lambda v \phi_1(\phi_1^2+\phi_2^2)
-{\lambda\over 4}(\phi_1^2+\phi_2^2)^2
\nonumber\\
&&
+ {\bar\psi}i\not \!\!\partial\psi +Gv{\bar\psi}\psi
+{e\over 2}{\bar\psi}\gamma_\mu\gamma_5\psi A^\mu
\nonumber\\
&&
+G{\bar\psi}\psi\phi_1
-iG{\bar\psi}\gamma_5\psi\phi_2
\nonumber\\
&&
+J_1 [-ec\phi_2] + J_2 ec(\phi_1+v)
+i{e\over 2}{\bar\eta}\gamma_5\psi c
+i{e\over 2}c{\bar\psi}\gamma_5 \eta
\Big ]
\label{41a}
\end{eqnarray}
The action is C invariant if the  
fields $\phi_2, A_\mu, i{\bar\psi}\gamma_5\psi, c,{\bar c}, J_2 $
are C odd and the fields $\phi_1, J_1 ,{\bar\psi}\psi$ are
C even. This invariance can be extended to $\eta, {\bar\eta}$ by
requiring
\begin{eqnarray}
&&
C\psi C^{-1} = B{\bar\psi}^T, \qquad C\eta C^{-1} = B{\bar\eta}^T \qquad
B^\dagger\gamma_\mu B = - \gamma_\mu^T
\nonumber\\
&&
B^2 =1 \quad B^* = B \quad B^T = - B \quad B^\dagger = B^{-1}
\label{41b}
\end{eqnarray}
Moreover we impose hermiticity for the low momentum expansion
of the vertex amplitude $\Gamma$ by requiring
\begin{eqnarray}
&&
c^\dagger =c \quad {\bar c}^\dagger = -{\bar c}
\nonumber\\
&&
 {\bar \eta}^\dagger = \gamma_0 \eta
\label{41c}
\end{eqnarray}
%
\section{ST transformation of counterterms}
\label{sec:ST}
The ST of the counterterms.
\par\noindent
The  scalar boson sectors.
\begin{eqnarray}
&&
S_0[\int d^4x \phi_1] = - e\int d^4x ( c \phi_2)
\nonumber \\
&&
S_0[\int d^4x \phi_1^2] = - 2e\int d^4x ( c \phi_1\phi_2)
\nonumber \\
&&
S_0[\int d^4x \phi_2^2] =  2 e\int d^4x (c \phi_2(\phi_1+v))
\nonumber \\
&&
S_0[\int d^4x \phi_1^3] = - 3e\int d^4x ( c \phi_1^2\phi_2)
\nonumber \\
&&
S_0[\int d^4x \phi_2^2 \phi_1] =   e\int d^4x c[ 2\phi_2(\phi_1+v)\phi_1
- \phi_2^3 ]
\nonumber \\
&&
S_0[\int d^4x \phi_1^4] = - 4 e\int d^4x (c\phi_1^3\phi_2)
\nonumber \\
&&
S_0[\int d^4x \phi_2^4] =  4 e\int d^4x (c \phi_2^3(\phi_1+v))
\nonumber \\
&&
S_0[\int d^4x \phi_2^2\phi_1^2] = e\int d^4x ( 2 c \phi_2\phi_1^2(\phi_1+v)
- 2c \phi_2^2\phi_2\phi_1)
\label{appb.41}
\end{eqnarray}
The kinetic boson sector
\begin{eqnarray}
&&
S_0[\int d^4x A^2] = - 2\int d^4x(c\partial_\mu A^\mu)
\nonumber \\
&&
S_0[\int d^4x\partial_\mu A^\mu\phi_2] =
\int d^4x c[\Box\phi_2 + e\partial_\mu A^\mu(\phi_1+v)]
\nonumber \\
&&
S_0[\int d^4x (\partial_\mu\phi_1)^2] = 2 e\int d^4x(c\Box\phi_1\phi_2)
\nonumber \\
&&
S_0[\int d^4x (\partial_\mu\phi_2)^2] = -2 e
\int d^4x(c\Box\phi_2(\phi_1+v))
\nonumber \\
&&
S_0[\int d^4x A^\mu\partial_\mu\phi_1\phi_2] =
\int d^4x c[-\Box\phi_1\phi_2 
- \partial_\mu\phi_1\partial^\mu\phi_2
\nonumber \\
&&
+e\partial_\mu A^\mu\phi_2^2
+ eA^\mu\phi_2\partial_\mu\phi_2
+ eA^\mu\partial_\mu\phi_1 (\phi_1+v)
]
\nonumber \\
&&
S_0[\int d^4x A^\mu\phi_1\partial_\mu\phi_2)] =
\int d^4x c[-\phi_1\Box\phi_2 
- \partial_\mu\phi_1\partial^\mu\phi_2
\nonumber \\
&&
- eA^\mu\phi_2\partial_\mu\phi_2
-e\partial_\mu A^\mu\phi_1(\phi_1+v)
- eA^\mu\partial_\mu\phi_1 (\phi_1+v)]
\nonumber \\
&&
S_0[\int d^4x A^2\phi_1] = \int d^4x c[-2\partial^\mu(
A_\mu\phi_1) - eA^2\phi_2]
\nonumber \\
&&
S_0[\int d^4x A^2\phi_1^2] = 2\int d^4x c[-\partial^\mu(
A_\mu\phi_1^2) - eA^2\phi_1\phi_2]
\nonumber \\
&&
S_0[\int d^4x A^2\phi_2^2] = 2\int d^4x c[-\partial^\mu(
A_\mu\phi_2^2) + eA^2\phi_2(\phi_1+v)]
\nonumber \\
&&
S_0[\int d^4x \partial_\mu A^\nu\partial^\mu A_\nu] =
2 \int d^4x c\Box\partial^\nu A_\nu
\nonumber \\
&&
S_0[\int d^4x \partial^\mu A_\mu\partial^\nu A_\nu] =
2 \int d^4x c\Box\partial^\nu A_\nu
\nonumber \\
&&
S_0[\int d^4x A^4] = - 4 \int d^4x c\partial^\mu(
A_\mu A^2)
\nonumber \\
&&
S_0[\int d^4x{\bar c}cA^2] = \int d^4x[{\cal F}cA^2
-2{\bar c}c(\partial^\mu c)A_\mu]
\label{appb.42}
\end{eqnarray}
Fermion sectors
\begin{eqnarray}
&&
S_0[\int d^4x {\bar\psi}i\gamma_\mu\partial^\mu\psi]
={e\over 2}\int d^4x c[{\bar\psi}\gamma_\mu\gamma_5\partial^\mu\psi
+(\partial^\mu{\bar\psi})\gamma_\mu\gamma_5\psi]
\nonumber\\
&&
S_0[\int d^4x {\bar\psi}\gamma_\mu\gamma_5 A^\mu\psi]
=- \int d^4x c[{\bar\psi}\gamma_\mu\gamma_5\partial^\mu\psi
+(\partial^\mu{\bar\psi})\gamma_\mu\gamma_5\psi]
\nonumber\\
&&
S_0[\int d^4x {\bar\psi}\psi]
=ie\int d^4x c[{\bar\psi}\gamma_5\psi]
\nonumber\\
&&
S_0[\int d^4x {\bar\psi}\psi\phi_1]
=e\int d^4x c[i{\bar\psi}\gamma_5\psi\phi_1-
{\bar\psi}\psi\phi_2]
\nonumber\\
&&
S_0[\int d^4x {\bar\psi}i\gamma_5\psi\phi_2]
=e\int d^4x c[-{\bar\psi}\psi\phi_2+i
{\bar\psi}\gamma_5\psi(\phi_1+v)]
\label{appb.43}
\end{eqnarray}
The ghost sector
\begin{eqnarray}
&&
S_0[\int d^4x {\bar c}c]= \int d^4x{\cal F}c
\nonumber\\
&&
S_0[\int d^4x{\bar c}\Box c]= \int d^4x{\cal F}\Box c
\nonumber\\
&&
S_0[\int d^4x{\bar c}c\phi_1]= \int d^4x{\cal F} c\phi_1
\nonumber\\
&&
S_0[\int d^4x{\bar c}c\phi_1^2]= \int d^4x{\cal F} c\phi_1^2
\nonumber\\
&&
S_0[\int d^4x{\bar c}c\phi_2^2]= \int d^4x{\cal F} c\phi_2^2
\label{appb.43a}
\end{eqnarray}
Fermion sources sector
\begin{eqnarray}
&&
S_0[\int d^4x~ {i\over 2} ({\bar\eta}\gamma_5\psi c
+c{\bar\psi}\gamma_5\eta)]
=c[{1\over 2}\partial^\mu({\bar\psi}\gamma_\mu\gamma_5\psi)
\nonumber\\ 
&&
+ iGv {\bar\psi}\gamma_5\psi 
+ iG {\bar\psi}\gamma_5\psi\phi_1 
+ G {\bar\psi}\psi\phi_2 ]
\label{appb.44}
\end{eqnarray}
\section{ST invariants}
\label{sec:STI}
We have two classes of ST invariants: the BRS invariants
where the sources do not intervene
\begin{eqnarray}
&&
{\cal I}_1 = \int d^4x (\phi_1^2 + \phi_2^2
+2 v\phi_1)
\nonumber\\ 
&&
{\cal I}_2 = \int d^4x (\phi_1^4 + \phi_2^4
+ 2\phi_1^2\phi_2^2 + 4 v\phi_1^3 + 4 v \phi_1\phi_2^2
+4v^2\phi_1^2)
\nonumber\\ 
&&
{\cal I}_3 = \int d^4x |D_\mu \phi|^2
\nonumber\\ 
&&
{\cal I}_4 = \int d^4x (F_{\mu\nu})^2
\nonumber\\ 
&&
{\cal I}_5 = \int d^4x {\bar \psi}i\gamma_\mu {\cal D}^\mu \psi
\nonumber\\ 
&&
{\cal I}_6 = \int d^4x {\bar \psi}[(\phi_1+v) -
i\gamma_5\phi_2]\psi
\nonumber\\ 
&&
{\cal I}_7  \equiv {\cal I}_7 = \int d^4x({1\over 2}{\cal F}^2 +
{\bar c}\delta_{\rm BRS}{\cal F})
\nonumber\\ 
&&
{\cal I}_{8}  \equiv {\cal I}_{8} = \int d^4x ({1\over 2} A^2 +{\bar c}c + 
{v\over\alpha}\phi_1 ) 
\label{appc.31}
\end{eqnarray}
and ST invariants with external sources:
\begin{eqnarray}
&&
{\cal I}_{9}  = 
\int d^4x [A^\mu\Gamma^{(0)}_{A^\mu}+
c\Gamma^{(0)}_c
+\alpha ({\cal F}\partial^\mu A_\mu - {\bar c}\Box c)]
\nonumber\\ 
&&
{\cal I}_{10}  = S_0(\int d^4x J_1) = 
\int d^4x \Gamma^{(0)}_{\phi_1}
\nonumber\\ 
&&
{\cal I}_{11}  = S_0(\int d^4x J_1\phi_1) = 
\int d^4x ( \phi_1\Gamma^{(0)}_{\phi_1}+ eJ_1 c\phi_2)
\label{appc.32}
\end{eqnarray}
There are other invariants which are linearly dependent from
the previous ones.
\begin{eqnarray}
&&
{\cal I}_{12} =
\int d^4x [\phi_2\Gamma^{(0)}_{\phi_2} - eJ_2 c(\phi_1+v)
+ev({\cal F}\phi_2 -e{\bar c}c(\phi_1+v))]
\nonumber\\ 
&&
= -\lambda v^2 {\cal I}_{1} - \lambda {\cal I}_{2} + 2{\cal I}_{3}
+ G {\cal I}_{6} - ev {\cal I}_{10} - {\cal I}_{11}
\nonumber\\ 
&&
{\cal I}_{13}  =\int d^4x (\Gamma^{(0)}
_\psi\psi -i{e\over 2}{\bar\eta}c\gamma_5\psi) = - {\cal I}_{5} -G {\cal I}_{6}
\nonumber\\ 
&&
{\cal I}_{14}  =\int d^4x ({\bar\psi}\Gamma^{(0)}_{\bar\psi} 
+ i{e\over 2}{\bar\psi}c\gamma_5\eta) = - {\cal I}_{13}
\label{appc.33}
\end{eqnarray}
\par
The coefficients of the invariant counterterms can
be fixed by choosing the normalization conditions
on some monomials. The following matrix provides
an example of the linear dependence of the ST invariants from
a set of monomials (for comparison an extra row
is added involving the fermi external source)  

\begin{eqnarray}
\left (
\begin{array}{llllllllll}
&v_1 &v_2 &v_3 &v_4 &v_5 &v_6 &v_9 &v_{10} &v_{11} \\
\phi_1  &2v &0 &0&0 &0&0 &0&-2\lambda v^2 &0  \\
\phi_2^2\phi_1  &0 &4v &0&0 &0&0 &0&-\lambda & -\lambda v \\
A^2\phi_1 &0&0 & e^2 v&0 &0&0 &2e^2 v&e^2 &e^2 v  \\
F_{\mu\nu}^2 &0&0 &0&1 &0&0 &0&0 &0 \\
{\bar \psi}\gamma_\mu\gamma_5\psi A^\mu  &0&0 &0&0 &{e\over 2}  &0
       &{e\over 2}  &0 &0 \\
i{\bar \psi}\gamma_5\psi \phi_2    &0&0 &0&0 &0&-1 &0&0 &0 \\
J_2c &0&0 &0&0 &0&0 &ev &e &0 \\
J_2c\phi_1  &0&0 &0&0 &0&0 &e &0 &e \\
J_1 c\phi_2  &0&0 &0&0 &0&0 &-e &0 &e \\
i{\bar \eta}\gamma_5\psi c &0&0 &0&0 &0&0 &{e\over 2} &0 &0
\end{array}
\right )
\label{hier.36b}
\end{eqnarray}
\section{Functional derivatives of $\Psi$}
\label{sec:psi}
\begin{enumerate}
\item
We project out the lowest powers of the momenta with $(1-t^3)$
and moreover we use the normalization conditions
\begin{eqnarray}
&&
\hat{\Psi}_{c \phi_2} = - e v\Xi^{(n)}_{\phi_2\phi_2}
\nonumber \\
&&
=({{(ev)^2}\over\alpha} -m_2^2)(t^3-t^1)\Gamma_{cJ_2}^{(n)}
+ t^3 \sum_{j=1}^{n-j}{{\rm I}\!\Gamma}^{(j)}_{cJ_2}{{\rm I}\!\Gamma}^{(n-j)}_{\phi_2^2}
=t^3 \sum_{j=1}^{n-j}{{\rm I}\!\Gamma}^{(j)}_{c J_2}{{\rm I}\!\Gamma}^{(n-j)}_{\phi_2^2}
\nonumber \\
\label{break.2}
\end{eqnarray}
In the Taylor expansion denoted by $t^3$, the odd-number
derivative of the vertex functions at zero momentum are zero, by
Lorentz covariance. Moreover the constant term is zero due to the
normalization condition ${{\rm I}\!\Gamma}^{(j)}_{cJ_2 }(0)=0$ thus 
only $\xi^{(n)}_{\partial_\mu\phi_2\partial^\mu\phi_2}$
can be non-zero. However we get
\begin{eqnarray}
- e v \Xi^{(n)}_{\phi_2\phi_2}
=\sum_{j=1}^{n-j}\left((t^2-t^1){{\rm I}\!\Gamma}^{(j)}_{cJ_2}
{{\rm I}\!\Gamma}^{(n-j)}_{\phi_2^2}(0)
+{{\rm I}\!\Gamma}^{(j)}_{cJ_2}(0)(t^2-t^1){{\rm I}\!\Gamma}^{(n-j)}_{\phi_2^2}
\right)=0
\nonumber \\
\label{break.3}
\end{eqnarray}
Thus finally ${\Xi}_{\phi_2\phi_2}$ is zero.
%
\item
Take the relevant functional derivatives of eq. (\ref{break.1}).
By using the normalization conditions and ${\Xi}_{\phi_2\phi_2}=0$
one gets
\begin{eqnarray}
&&
{\hat\Psi}_{c\phi_2(p)\phi_1(q)}^{(n)} =
-[i(p+q)^\mu \Xi_{A^\mu\phi_2(p)\phi_1(q)}^{(n)} 
-e\Xi^{(n)}_{\phi_1\phi_1(q)} ]
=
\nonumber\\&&
=-m_1^2 (t^2-t^1)\Gamma_{cJ_1(q)\phi_2(p)}^{(n)} + ev(t^2-t^1)
\Gamma^{(n)}_{\phi_2\phi_2(p)\phi_1(q)}
- 2 \lambda v (t^2-t^1)\Gamma_{cJ_2(p+q)}^{(n)} 
\label{break.5}
\end{eqnarray}
The lower order contributions are zero by the normalization conditions.
Since $\Psi^{(n)}_{c\phi_2\phi_1}$ contains only terms quadratic in
the momenta, then there is no counterterm as $\int d^4x\phi_1^2$.
It should be reminded that we have already chosen 
$\Xi_{\phi_2^2\phi_1}=0$. 
%
\item
\begin{eqnarray}
{\hat\Psi}^{(n)}_{c\phi_2\phi_1^2}
=- \left[ -e\Xi^{(n)}_{\phi_1^3}
+ ev
\Xi^{(n)}_{\phi_2^2\phi_1^2} \right]
= - 2 m_1^2 \Gamma_{cJ_1\phi_2\phi_1}^{(n)}(0)
\label{break.7}
\end{eqnarray}
\par\noindent
%
%
%
\item
\begin{eqnarray}
&&{\hat\Psi}^{(n)}_{c\phi_2\phi_1^3}
=-[-e\Xi^{(n)}_{\phi_1^4} 
+ 3e\Xi^{(n)}_{\phi_2^2\phi_1^2}]
\nonumber\\&&
=-18\lambda v\Gamma_{cJ_1\phi_2\phi_1}^{(n)}(0)
-3m_1^2 \Gamma_{cJ_1\phi_2\phi_1^2}^{(n)}(0)
+ev\Gamma^{(n)}_{\phi^2_2 \phi_1^3}(0)
-6\lambda v
\Gamma_{cJ_2\phi_1^2}^{(n)}(0)
\nonumber\\&&
 + 3 \sum_{j=1}^{n-1} \left( {{\rm I}\!\Gamma}_{\phi^3}^{(n-j)}(0) 
{{\rm I}\!\Gamma}_{cJ_1\phi_1\phi_2}^{(j)} \right)
\label{break.9}
\end{eqnarray}
\par\noindent%
%
\item
\begin{eqnarray}
{\hat\Psi}^{(n)}_{c\phi_2^3} = 0
\label{break.11}
\end{eqnarray}
%
\item
\begin{eqnarray}
&&
{\hat\Psi}^{(n)}_{c\phi_2^3\phi_1}=
-[-3e\Xi^{(n)}_{\phi_1^2\phi_2^2} 
+ e\Xi^{(n)}_{\phi_2^4}]
\nonumber\\&&
= 
-6 \lambda v\Gamma_{c J_1\phi_2\phi_1}^{(n)}(0)
+ev\Gamma^{(n)}_{\phi_2^4\phi_1}(0)
-m_1^2\Gamma_{c J_1\phi_2^3}^{(n)}(0)
-6\lambda v\Gamma_{c J_2\phi_2^2}^{(n)}(0)
\label{break.13}
\end{eqnarray}
\par\noindent
%
\item
\begin{eqnarray}
{\hat\Psi}^{(n)}_{cA_\nu}
= - i p^\mu \Xi^{(n)}_{A^\mu A_\nu}
=ev(t_p^3-t_p^2)\Gamma^{(n)}_{\phi_2 A_\nu}
- i e v p^\nu(t_p^2-t_p^1)\Gamma_{cJ_2}^{(n)}
\label{break.15}
\end{eqnarray}
Then there is no contribution to $A^2$ and to the 
transverse part of $A_\mu$, i.e. $\int d^4x F_{\mu\nu}^2$. 
Only to $(\partial_\mu A^\mu)^2$.
%
\item
\begin{eqnarray}
&&{\hat\Psi}^{(n)}_{cA_\nu(p)\phi_1(q)}=
- \left[ i(p+q)^\mu \Xi^{(n)}_{A^\mu A_\nu(p)\phi_1(q)} 
+ e v \Xi^{(n)}_{\phi_2 A_\nu(p)\phi_1(q)} \right]
\nonumber\\&& 
= - m_1^2 (t^2-t^0) \Gamma^{(n)}_{cJ_1(q) A_\nu(p)}
+ev(t^2-t^1)\Gamma^{(n)}_{\phi_2 A_\nu(p)\phi_1(q)} 
\label{break.17}
\end{eqnarray}
The last term is zero because of covariance. 
%
\item
\begin{eqnarray}
&&{\hat\Psi}^{(n)}_{cA_\nu(p)\phi_1(q_1)\phi_1(q_2)} 
\nonumber\\&&
= - \left[
i(p+q_1+q_2)^\mu \Xi^{(n)}_{A^\mu A_\nu(p)\phi_1(q_1)\phi_1(q_2) } 
+ e\Xi^{(n)}_{\phi_2 A_\nu(p)\phi_1(q_1)} 
+ e\Xi^{(n)}_{\phi_2 A_\nu(p)\phi_1(q_2)} \right]
=
\nonumber\\&&  
- m_1^2( t^1 \Gamma^{(n)}_{cJ_1 A_\nu(p)\phi_1(q_1)}
+t^1 \Gamma^{(n)}_{J_1c A_\nu(p)\phi_1(q_2)})
- 6 \lambda v (t^1-t^0) 
\Gamma^{(n)}_{cJ_1(q_1+q_2) A_\nu(p)}
\nonumber\\&&  
+ev(t^1-t^0)\Gamma^{(n)}_{\phi_2 A_\nu(p)\phi_1(q_1)\phi_1(q_2)} 
-iev p^\nu \Gamma^{(n)}_{cJ_2(p)\phi_1(q_1)\phi_1(q_2)}
+\sum_{j=1}^{n-1}\big({{\rm I}\!\Gamma}_{\phi_1^3 }^{(n-j)}(0)
t^1{{\rm I}\!\Gamma}_{c J_1 A_\nu}^{(j)}
\big) 
\nonumber\\
\label{break.19}
\end{eqnarray}
Notice that the breaking-term $\Gamma^{(0)}_{\phi_2 A_\nu\phi_1}
(t^1-t^0)\Gamma^{(n)}_{c J_2\phi_1}$ is zero and 
therefore it has been omitted. 
%
\item
\begin{eqnarray}
&&{\hat\Psi}^{(n)}_{cA_\nu(p)\phi_2(q_1)\phi_2(q_2)}
\nonumber\\&& 
= - \left[ 
i(p+q_1+q_2)^\mu \Xi^{(n)}_{A^\mu A_\nu(p)\phi_2(q_1)\phi_2(q_2)}
-e\Xi^{(n)}_{\phi_1 A_\nu(p)\phi_2(q_1)} 
-e\Xi^{(n)}_{\phi_1 A_\nu(p)\phi_2(q_2)} \right]
\nonumber\\&& 
= - 2 \lambda v (t^1-t^0)
\Gamma_{c J_1(q_1+q_2)A_\nu(p)}^{(n)}
+ev(t^1-t^0)\Gamma^{(n)}_{\phi_2 A_\nu(p)\phi_2(q_1)\phi_2(q_2)}
+ i p^\nu v \Gamma^{(n)}_{c J_2 \phi_2^2}(0)
\nonumber\\
\label{break.22}
\end{eqnarray}
%
%
\item
\begin{eqnarray}
&&{\hat\Psi}^{(n)}_{cA_\mu(p_1) A_\nu(p_2)\phi_2(q)}= - \left[ 
-e\Xi^{(n)}_{\phi_1A_\mu(p_1) A_\nu(p_2)} 
+ ev \Xi^{(n)}_{\phi_2A_\mu A_\nu\phi_2(q)} \right]
\nonumber\\&& 
= 
ev(t^1-t^0)\Gamma_{\phi_2 A_\mu(p_1) A_\nu(p_2)\phi_2(q)}^{(n)}
+ 2 e v^2 (t^1-t^0) g^{\mu \nu}
\Gamma_{cJ_1(p_1+p_2)\phi_2(q)}^{(n)}
iev(p^{\mu}_1\Gamma_{c J_1(p_1)A_\nu(p_2)\phi_2(q)}^{(n)}(0) 
\nonumber\\&& 
+p^{\nu}_2\Gamma_{c J_1(p_2)A_\mu(p_1)\phi_2(q)}^{(n)}(0) )
+\sum_{j=1}^{n-1}\big(
{{\rm I}\!\Gamma}_{\phi_1 A_\mu A_\nu }^{(n-j)}(0) 
t^1{{\rm I}\!\Gamma}_{c J_1(p_1+p_2)\phi_2(q)}^{(j)}
\big) 
\nonumber\\
\label{break.24}
\end{eqnarray}
>From the Lorentz structure we see that all terms are zero and
therefore 
\begin{eqnarray}
{\hat\Psi}^{(n)}_{cA_\mu(p_1) A_\nu(p_2)\phi_2(q)}=
 e\Xi^{(n)}_{\phi_1A_\mu(p_1) A_\nu(p_2)} 
- ev \Xi^{(n)}_{\phi_2A_\mu A_\nu\phi_2(q)}=0
\label{break.24p}
\end{eqnarray}
%
\item
\begin{eqnarray}
&&{\hat\Psi}^{(n)}_{cA_\mu A_\nu\phi_2\phi_1}
=- \left[ 
-e\Xi^{(n)}_{\phi_1A_\mu A_\nu\phi_1} 
+ e\Xi^{(n)}_{\phi_2A_\mu A_\nu\phi_2} \right]
\nonumber\\&& 
= ev\Gamma_{\phi_2A_\mu A_\nu\phi_2\phi_1}^{(n)}(0)
-m_1^2 \Gamma_{c J_1 A_\mu A_\nu\phi_2}^{(n)}(0)
+ 2 e v^2 g^{\mu\nu}\Gamma_{c J_1 \phi_2\phi_1}^{(n)}(0)
\nonumber\\&&  
- 2 \lambda v \Gamma_{c J_2 A_\mu A_\nu}^{(n)}(0)
+\sum_{j=1}^{n-1}\big({{\rm I}\!\Gamma}_{cJ_1\phi_2\phi_1}^{(j)}
{{\rm I}\!\Gamma}_{\phi_1 A_\mu A_\nu }^{(n-j)}
\big) (0)
\label{break.26}
\end{eqnarray}
%
%
\item
\begin{eqnarray}
&&{\hat\Psi}^{(n)}_{cA_\nu A_\rho A_\sigma}
= - \left[ 
i(p_1+p_2+p_3)_\mu \Xi^{(n)}_{A_\mu A_\nu A_\rho A_\sigma} \right]
\nonumber\\&& 
=
ev t^1 \Gamma^{(n)}_{\phi_2 A_\nu A_\rho A_\sigma} 
+2 e v^2 g^{\rho\sigma} (t^1-t^0)
\Gamma_{cJ_1(p_2+p_3)A_\nu(p_1)}^{(n)}
\nonumber\\&& 
+ 2 e v^2 g^{\nu\sigma} (t^1-t^0)
\Gamma_{cJ_1(p_1+p_3)A_\rho(p_2)}^{(n)}
+ 2 e v^2 g^{\nu\rho}(t^1-t^0)
\Gamma_{cJ_1(p_1+p_2)A_\sigma(p_3)}^{(n)}
\nonumber\\&& 
+ie v p^{\nu}_1\Gamma^{(n)}_{cJ_2(p_1) A_\rho A_\sigma}(0)
+ie v p^{\rho}_2\Gamma^{(n)}_{cJ_2(p_2) A_\mu A_\sigma}(0)
+ie v p^{\sigma}_3\Gamma^{(n)}_{cJ_2(p_3) A_\mu A_\rho}(0)
\nonumber\\&& 
+ \sum_{j=1}^{n-1}\big(t^1{{\rm I}\!\Gamma}_{cJ_1A_\rho}^{(j)}
\big){{\rm I}\!\Gamma}_{\phi_1 A_\sigma A_\nu }^{(n-j)}(0)
+ \sum_{j=1}^{n-1}\big(t^1{{\rm I}\!\Gamma}_{cJ_1A_\nu}^{(j)}
\big){{\rm I}\!\Gamma}_{\phi_1 A_\sigma A_\rho }^{(n-j)}(0)
+ \sum_{j=1}^{n-1}\big(t^1{{\rm I}\!\Gamma}_{cJ_1A_\sigma}^{(j)}
\big){{\rm I}\!\Gamma}_{\phi_1 A_\rho A_\nu }^{(n-j)}(0)
\nonumber\\ 
\label{break.28}
\end{eqnarray}
\item
\begin{eqnarray}
&&{\hat\Psi}^{(n)}_{c{\bar\psi}(p_1)\psi(p_2)}
= - \left[ 
i{e\over 2}\gamma_5\Xi^{(n)}_{\psi(p_2){\bar\psi}}
+i{e\over 2}\Xi^{(n)}_{\psi{\bar\psi}(p_1)}\gamma_5 
-Gv\Xi^{(n)}_{{\bar\eta}(p_1)c\psi(p_2)}
-Gv\Xi^{(n)}_{\eta(p_2){\bar\psi}(p_1)c}
\right]
\nonumber\\&& 
=
-ev (t^1-t^0) \Gamma^{(n)}_{\phi_2\psi(p_2){\bar\psi}(p_1) }
- G v (t^1-t^0)
\Gamma_{{\bar\eta}(p_1)c\psi(p_2)}^{(n)}
- G v (t^1-t^0)\Gamma_{\eta(p_2){\bar\psi}(p_1)c}^{(n)}
\nonumber\\&& 
-t^1\sum_{j=1}^{n-1}\big(
{{\rm I}\!\Gamma}_{\eta(p_2){\bar\psi}(p_1)c}^{(j)}
{{\rm I}\!\Gamma}^{(n-j)}_{\psi(p_2){\bar\psi}}
+{{\rm I}\!\Gamma}^{(j)}_{\psi{\bar\psi}(p_1)}
{{\rm I}\!\Gamma}_{{\bar\eta}(p_1)c\psi(p_2)}^{(n-j)}
\big)
\label{break.29}
\end{eqnarray}
\item
\begin{eqnarray}
&&{\hat\Psi}^{(n)}_{c{\bar\psi}(p_1)\psi(p_2)\phi_1(q)}
\nonumber\\&& 
= - \left[ 
i{e\over 2}\gamma_5\Xi^{(n)}_{{\bar\psi}\psi(p_2)\phi_1(q)}
+i{e\over 2}\Xi^{(n)}_{\psi{\bar\psi}(p_1)\phi_1(q)}\gamma_5 
-e \Xi^{(n)}_{\psi(p_2){\bar\psi}(p_1)\phi_2(q)}
-G\Xi^{(n)}_{{\bar\eta}(p_1)c\psi(p_2)}
-G\Xi^{(n)}_{\eta(p_2){\bar\psi}(p_1)c}
\right]
\nonumber\\&& 
=
-ev  \Gamma^{(n)}_{\phi_2{\bar\psi}(p_1)\psi(p_2)\phi_1(q) }
-m_1^2\Gamma^{(n)}_{c J_1{\bar\psi}(p_1)\psi(p_2)}
- G v 
\Gamma_{{\bar\eta}(p_1)c\psi(p_2)\phi_1(q)}^{(n)}
- G v \Gamma_{\eta(p_2){\bar\psi}(p_1)c\phi_1(q)}^{(n)}
\nonumber\\&& 
-t^0\sum_{j=1}^{n-1}\big(
{{\rm I}\!\Gamma}_{\eta(p_2){\bar\psi}(p_1)c}^{(j)}
{{\rm I}\!\Gamma}^{(n-j)}_{\psi(p_2){\bar\psi}\phi_1(q)}
+{{\rm I}\!\Gamma}^{(j)}_{\psi{\bar\psi}(p_1)\phi_1(q)}
{{\rm I}\!\Gamma}_{{\bar\eta}(p_1)c\psi(p_2)}^{(n-j)}
\big)
\nonumber\\&& 
-t^0\sum_{j=1}^{n-1}\big(
{{\rm I}\!\Gamma}_{\eta(p_2)c{\bar\psi}(p_1)\phi_1(q)}^{(j)}
{{\rm I}\!\Gamma}^{(n-j)}_{\psi(p_2){\bar\psi}}
+{{\rm I}\!\Gamma}^{(j)}_{\psi{\bar\psi}(p_1)}
{{\rm I}\!\Gamma}_{{\bar\eta}(p_1)c\psi(p_2)\phi_1(q)}^{(n-j)}
\big)
\label{break.30}
\end{eqnarray}
\item
\begin{eqnarray}
&&{\hat\Psi}^{(n)}_{c{\bar\psi}(p_1)\psi(p_2)\phi_2(q)}
= - \left[ 
+e\Xi^{(n)}_{\psi(p_2){\bar\psi}\phi_1(q)} 
-iG\Xi^{(n)}_{{\bar\eta}(p_1)c(-p_1-p_2-q)\psi(p_2)}
-iG\Xi^{(n)}_{\eta(p_2){\bar\psi}(p_1)c(-p_1-p_2-q)}
\right]
\nonumber\\&& 
=
-ev  \Gamma^{(n)}_{\phi_2\psi(p_2){\bar\psi}(p_1)\phi_2(q) }
- G v 
\Gamma_{{\bar\eta}(p_1)c\psi(p_2)\phi_2(q)}^{(n)}
- G v \Gamma_{\eta(p_2){\bar\psi}(p_1)c\phi_2(q)}^{(n)}
\nonumber\\&& 
-t^0\sum_{j=1}^{n-1}\big(
{{\rm I}\!\Gamma}_{\eta(p_2){\bar\psi}(p_1)c\phi_2(q)}^{(j)}
{{\rm I}\!\Gamma}^{(n-j)}_{\psi(p_2){\bar\psi}}
+{{\rm I}\!\Gamma}^{(j)}_{\psi{\bar\psi}(p_1)}
{{\rm I}\!\Gamma}_{{\bar\eta}(p_1)c\psi(p_2)\phi_2(q)}^{(n-j)}
\big)
\label{break.31}
\end{eqnarray}
$\hat{\Psi}_{c \bar{\psi} \gamma_\mu \psi A^{\mu}}=0$. In fact 
the only possible counterterm is 
${\bar\psi}\gamma_\mu\gamma_5\psi A^\mu$ and this is excluded by the
normalization conditions.
\end{enumerate}
%
\par
The above analysis shows that at every order the following
counterterms are absent to all orders.
\begin{eqnarray}
\int d^4 x \phi_1^2 \quad
\int d^4 x \phi_2^2 \quad
\int d^4 x (\partial_\mu\phi_2)^2 \quad
\int d^4 x A_\mu^2.
\label{break.40}
\end{eqnarray}
%

%
\end{document}